\input epsf

\newfam\msbfam
\font\twlmsb=msbm10 at 12pt
\font\eightmsb=msbm10 at 8pt
\font\sixmsb=msbm10 at 6pt
\textfont\msbfam=\twlmsb
\scriptfont\msbfam=\eightmsb
\scriptscriptfont\msbfam=\sixmsb
\def\cj{\fam\msbfam}

\def\R{{\cj R}}

\centerline{\bf RINDLER SPACE AND UNRUH EFFECT} 

\

\

\centerline{M. Socolovsky}

\

\centerline{\it  Instituto de Ciencias Nucleares, Universidad Nacional Aut\'onoma de M\'exico}
\centerline{\it Circuito Exterior, Ciudad Universitaria, 04510, M\'exico D. F., M\'exico} 

\

{\bf Abstract.} {\it We review the geometry of the Rindler space induced by hyperbolic motion in special relativity, and its applications to the calculation of the Unruh effect in flat spacetime, and to the Hawking temperature of the Schwarzschild black hole.}

\

{\bf I. Introduction.}

\

Hyperbolic motion in Minkowski spacetime (Landau and Lifshitz, 1975), that is, classical motion of a relativistic particle with constant proper acceleration, plays an important role in the understanding of motion in the presence of horizons and, most important, in the understanding of relevant phenomena such as Hawking radiation in black holes and the corresponding temperature (Hawking, 1974, 1975). It was precisely investigations on this subject, that led Unruh (Unruh, 1976) to discover a thermal effect in the vacuum of a quantum field in Minkowski space when this vacuum is observed from a uniformly accelerated frame which, in terms of the adapted Rindler coordinates (Rindler, 1966), has a fixed value for the spatial coordinate and an evolving temporal coordinate. 

\

It is the purpose of the present article to review, to some extend, part of this subject, in particular with some insights into the geometry of the Rindler space and its three additional wedges, its Penrose diagram, here considered as a space in itself, its maximal analytic extension to Minkowski spacetime, its role as the ``theater" where the Unruh effect occurs, and its application to the determination of the Hawking temperature of the evaporating Schwarzschild black hole.

\

In subsection II.1. we define light cone coordinates $u$ and $v$ in Minkowski space. Restricting to 1+1 dimensions ($Mink^2$), in subsection II.2. a detailed definition of hyperbolic motion is given and is shown how the concepts of horizons, invisible regions, and the Rindler wedge $R$ appear. The description of the particle motion in terms of the proper time $\tau$ and proper acceleration $\alpha$ is given in subsection II.3., while Rindler coordinates $\xi$ (spatial) and $\lambda$ (temporal) are defined in subsection II.4. Here it is shown how $\alpha$ is written in terms of $\xi$ and a constant $a$ which provides the units of acceleration, and it is exhibited the relation between $\alpha$, $a$, $\lambda$ and $\tau$, which is crucial for identifying the correct Unruh temperature ${\cal T}$ in subsection III.2. In terms of dimensionless Rindler coordinates $\rho$ and $\eta$ defined in subsection II.5., we study timelike geodesic motion in $R$ showing its geodesic incompleteness. In subsection II.6. we develop the steps to construct $Mink^2$ from $R$ as its maximal analytic extension, while in subsection II.7., left ($L$), future ($F$) and past ($P$) Rindler wedges are constructed, and a brief discussion of the discrete symmetries of parity, time reversal, and charge conjugation between the accelerated trajectories in $R$ and $L$ is presented. Future directed (in Rindler time $\lambda$) trajectories are interpreted as particles in $R$ and antiparticles in $L$. Light cone Rindler coordinates $\bar{u}$ and $\bar{v}$ are defined in subsection II.8., together with positive frequency right and left moving solutions of the Klein-Gordon equation, in preparation for the calculation of the Unruh effect in section {\bf III}. The final subsection II.9. of section {\bf II}, gives a detailed presentation of the construction of the Penrose diagrams, here considered as spaces in its own right (compact pseudo-Riemannian manifolds with boundary) of the Minkowski and Rindler spaces. The Unruh effect is calculated in two different ways in section {\bf III}. In III.1., following the original lines of thought (Unruh, 1976), through the application of the Bogoliubov transformation, and in III.2. the direct calculation of Lee (Lee, 1986) is reproduced in detail. In particular the discussion emphasizes how the proper acceleration $\alpha$ (and not $a$) appears naturally in the final result for the Unruh temperature (using the relation of equation (19)), without explicitly appealing to the red shift between different accelerated trajectories (Carroll, 2004; Birrell and Davies, 1982). Finally, in section {\bf IV}, the approximation of the Schwarzschild metric in the neighborhood of the horizon of the corresponding black hole in subsection IV.2., gives rise to a 2-dimensional Rindler space (Raine and Thomas, 2010). This allows to apply the results of section {\bf III} to obtain the Hawking temperature of the thermal radiation from the black hole in terms of its surface gravity, which is defined in subsection IV.1. A different approach to this problem, based on a global embedding in a higher dimensional Minkowski spacetime (GEMS), was developed by Deser and Levin (Deser and Levin, 1999). 

\

{\bf II. Hyperbolic motion. Rindler space. Four wedges of Minkowski spacetime.}

\

{\it II.1. Minkowski spacetime}

\

Let $x^\mu=(x^0,\vec{x})=(ct,x,y,z)=(ct,x^1,x^2,x^3)$ be the global chart of Minkowski spacetime. The metric is given by $$ds^2=\eta_{\mu\nu}dx^\mu dx^\nu =c^2dt^2-\vert d\vec{x}\vert ^2, \eqno{(1)}$$ with $$\eta_{\mu\nu}=diag(1,-1,-1,-1). \eqno{(2)}$$ For later use we define the light cone coordinates $$\pmatrix{u \cr v \cr}=\pmatrix{1 & -1 \cr 1 & 1 \cr} \pmatrix{ct \cr x} \eqno{(3)}$$ i.e. $u=ct-x$ and $v=ct+x$, which lead to $$ds^2=dudv-\vert d\vec{x}_{\perp}\vert ^2, \ \ \vec{x}_{\perp}=(x^2,x^3) \eqno{(4)}$$ i.e. to the metric $$\eta_{LC}=\pmatrix{0 & {{1}\over{2}} & 0 & 0 \cr {{1}\over{2}} & 0 &0 & 0 \cr 0 & 0 & -1 & 0 \cr 0 & 0 & 0 & -1}. \eqno{(5)}$$ $u=const.$ implies $ct=x+const.$ and $v=const.^\prime$ implies $ct=-x+const.^\prime.$ Also, with the metric $(g)=\eta_{LC}$, $g_{uu}=g(\partial_u,\partial_u)=\vert\vert{{\partial}\over{\partial u}}\vert\vert^2=0$ and $g_{vv}=g(\partial_v,\partial_v)=\vert\vert{{\partial}\over{\partial v}}\vert\vert^2=0$ i.e. ${{\partial}\over{\partial u}}$ and ${{\partial}\over {\partial v}}$ are null vectors. They are not orthogonal since $g_{uv}=g_{vu}=g(\partial_u,\partial_v)={{1}\over{2}}.$ 

\

$u=v=+\infty (-\infty)$ define the future (past) null infinity. (See Figure 1.)

\

\centerline{\epsfxsize=36ex\epsfbox{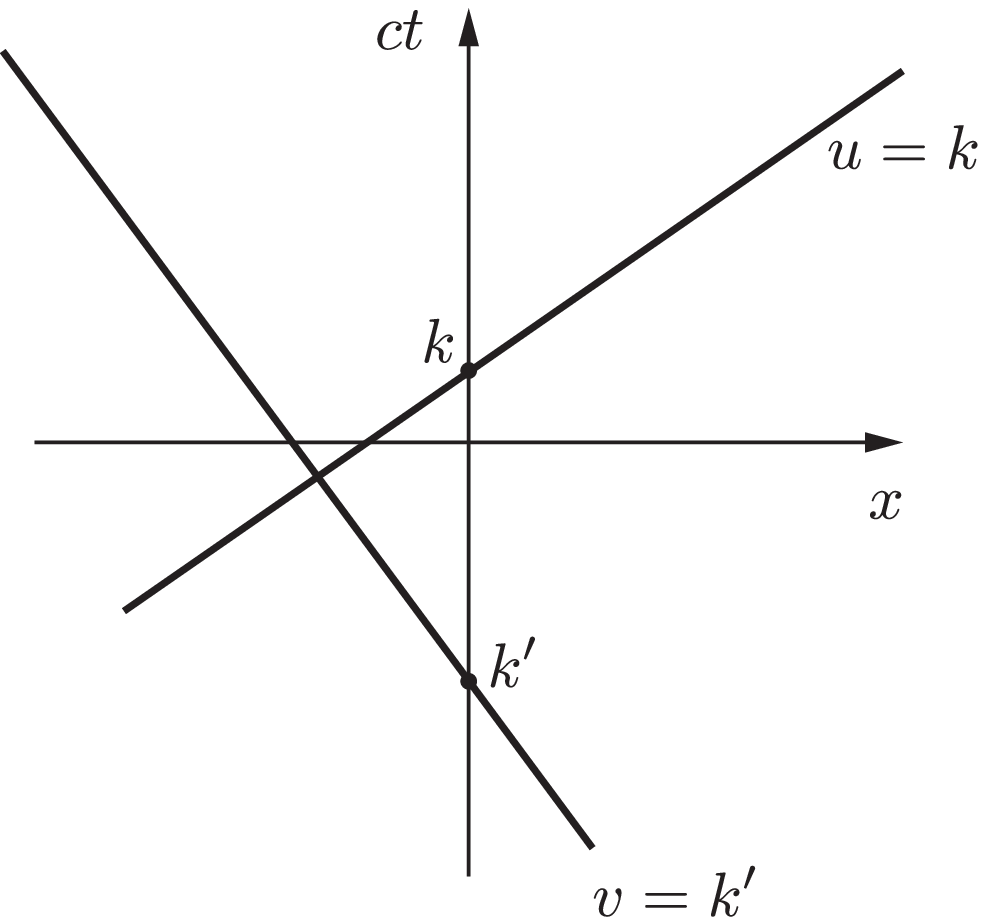}}

\

\centerline{Figure 1. Null (light cone coordinates) in Minkowski spacetime}

\

{\it II.2. Hyperbolic motion}

\

Let $u^\mu=(c\gamma,\gamma\vec{v})$ be be the 4-velocity of a particle, with $\gamma=(1-v^2/c^2)^{-{{1}\over{2}}}$ and $\vec{v}$ the ordinary 3-velocity ($[v]=[L]/[T]$). The 4-momentum of the particle is $p^\mu=mu^\mu=(E/c,\vec{p})$ where $E=\gamma mc^2$, $\vec{p}=\gamma m\vec{v}$, and $m$ is the mass of the particle. The relation between the proper time and $t$ is $dt=\gamma d\tau$, so for the 4-acceleration we have $$\alpha^\mu={{du^\mu}\over{d\tau}}=\gamma{{du^\mu}\over{dt}}=\gamma(c{{d\gamma}\over{dt}},{{d(\gamma\vec{v})}\over{dt}})=(\alpha^0,\vec{\alpha}) \eqno{(6)}$$ where $$\alpha^0=c\gamma\dot{\gamma}, \ \ \ \vec{\alpha}=\gamma{{d(\gamma\vec{v})}\over{dt}}.\eqno{(7)}$$ $[\alpha^\mu]=[L]/[T]^2$. In the instantaneous rest system of the particle, $\gamma=1$, so the {\it proper acceleration} is $$\alpha^\mu\vert_{RS}=(c\dot{\gamma}\vert_{\vec{v}=0},{{d(\gamma\vec{v})}\over{dt}}\vert_{\vec{v}=0})=(0,{{d^2\vec{x}}\over{dt^2}})\eqno{(8)}$$ since $\dot{\gamma}\vert_{\vec{v}=\vec{0}}=-{{1}\over{2}}(1-v^2/c^2)^{-{{3}\over{2}}}(-2\vec{v}\cdot {{d\vec{v}}\over{dt}})\vert_{\vec{v}=\vec{0}}=0$ and ${{d(\gamma\vec{v})}\over{dt}}\vert_{\vec{v}=\vec{0}}=(\dot{\gamma}\vec{v}+\gamma {{d\vec{v}}\over{dt}})\vert_{\vec{v}=\vec{0}}={{d\vec{v}}\over{dt}}=\vec{\alpha}\vert_{RS}={{d^2\vec{x}}\over{dt^2}}.$ This shows that $\alpha^\mu$ is spacelike since $\eta_{\mu\nu}\alpha^\mu\alpha^\nu=-\vec{\alpha}^2<0$. {\it Hyperbolic motion} is defined as that in which $\vec{\alpha}\vert_{RS}$ is constant. (In particular, if the particle has electric charge $q$, a constant electric field $\vec{\epsilon}$ produces hyperbolic motion since for the force one has $\vec{F}\vert_{RS}=q\vec{\epsilon}=m\vec{\alpha}\vert_{RS}$.)

\

Let us restrict our analysis to 1+1 dimensions i.e. to the $(t,x)$ plane of Minkowski spacetime, $Mink^2$. From $\eta_{\mu\nu}\alpha^\mu\alpha^\nu=(\alpha^0)^2-(\alpha^1)^2=({{du^0}\over{d\tau}})^2-({{du^1}\over{d\tau}})^2=\gamma^2(({{du^0}\over{dt}})^2-({{du^1}\over{dt}})^2)=-\alpha^2$ one obtains the differential equation $$\alpha={{d(\gamma v)}\over{dt}}={{d(v(1-v^2/c^2)^{-{{1}\over{2}}})}\over{dt}}=const. \eqno{(9)}$$ (Without loss of generality we take $\alpha>0$.) Then $\alpha dt=d(v/(1-v^2/c^2)^{{{1}\over{2}}})$; if $v(t_0=0)=0$, then $\alpha t=v(t)/(1-v(t)^2/c^2)^{{{1}\over{2}}}$ i.e. $$v(t)={{\alpha t}\over{(1+(\alpha t/c)^2)^{{1}\over{2}}}}. \eqno{(10)}$$ Notice that $v(t)\to c_{-}$ as $t\to +\infty$, and that for small values of $t$, $v\simeq\alpha t$. Also, ${{dx(t)}\over{dt}}=v(t)$, which implies $$x(t)=\alpha\int_0^t{{dt^\prime t^\prime}\over{\sqrt{1+(\alpha t^\prime /c)^2}}}=(c^2/\alpha)\sqrt{1+(\alpha t/c)^2} \eqno{(11)}$$ with $x(0)=c^2\alpha^{-1}$. For small $t$, $x(t)=(c^2/\alpha)+{{1}\over{2}}\alpha t^2$, while for large $\vert t\vert$, $x(t)\simeq c\vert t\vert$ i.e. $x(t)\to c\vert t\vert_+$ as $t\to \pm\infty$. The motion is represented by the hyperbola $$x^2-c^2t^2=(c^2/\alpha)^{2}\eqno{(12)}$$ which is plotted in Figure 2. It begins at $x=+\infty$ at $t=-\infty$ (past null infinity $u=-\infty$), comes to stop at $t=0$ for $x=(c^2/\alpha)$ and comes back to $x=+\infty$ for $t=+\infty$ (future null infinity $v=+\infty$). As $\alpha$ grows the hyperbola approaches the light lines $x=ct$ for $t>0$ and $x=-ct$ for $t<0$. The region interior to these lines is called the {\it Rindler right wedge} $R$ of $Mink^2$ or, more simply, {\it Rindler space}. It is covered by the infinite set of uniformely accelerated motions with $\alpha^{-1}\in (0,+\infty)$. It is clear that no signal can arrive to any point $p$ of the trajectory of the particle from points above or at the light line $x=ct$ which is therefore an {\it horizon} for the motion. The region $ct\geq x$ is {\it invisible} for the accelerated particle. At the same time, no information can reach from $R$ to points with $ct<-x$. Then the segments $x=ct$, $t>0$ and $x=-ct$, $t<0$ are called the {\it future} and {\it past} horizons of $R$. Finally, as $\alpha\to +\infty$ the hyperbola degenerates into the above mentioned lines, so  for light proper acceleration is infinite.

\

\centerline{\epsfxsize=50ex\epsfbox{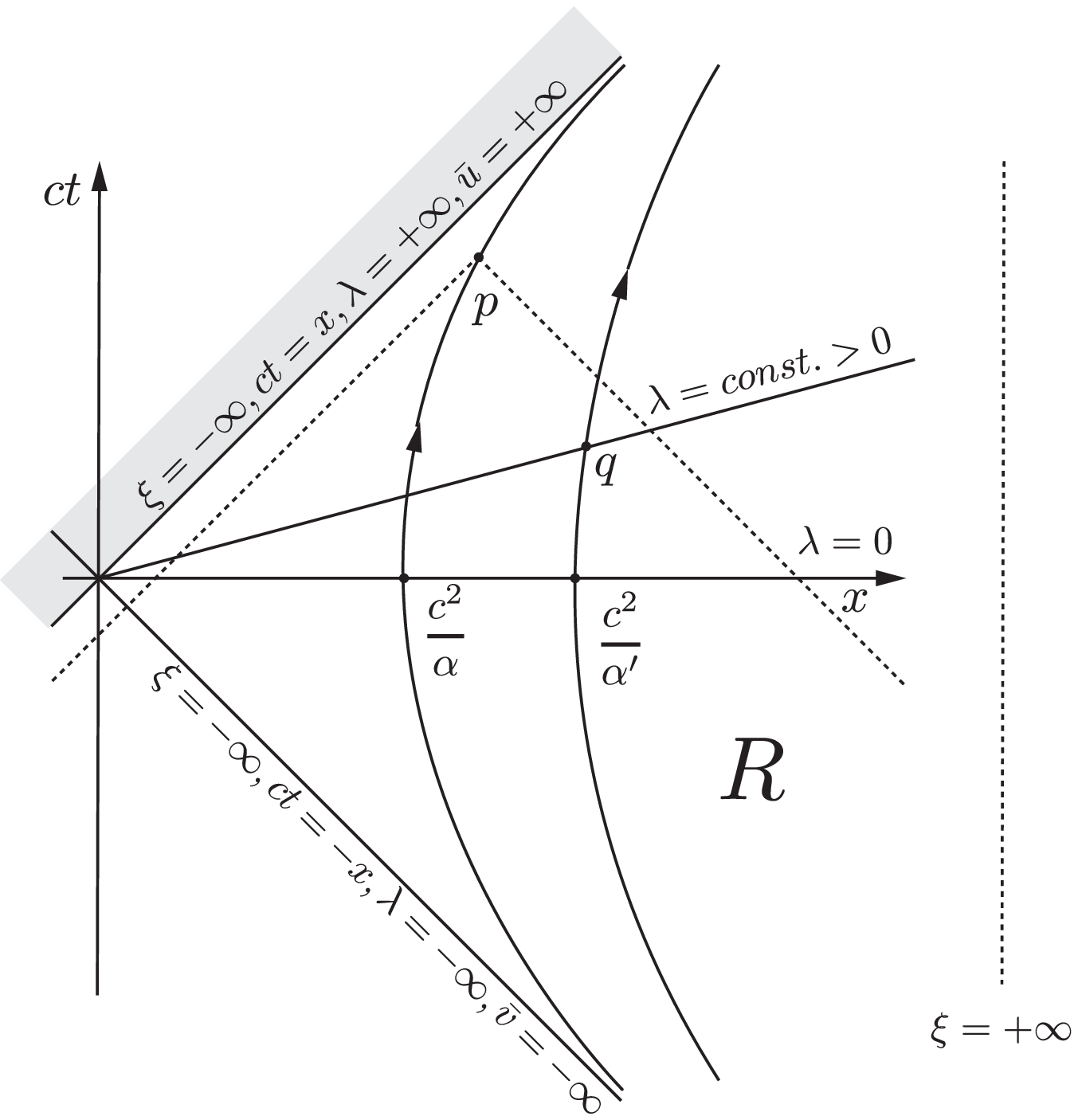}}

\

\centerline{Figure 2. Hyperbolic motion in the right Rindler wedge}

\

{\it II.3. Proper time notation}

\

For the proper time of the particle one has $\tau=\int_0^tdt^\prime\sqrt{1-v(t^\prime)^2/c^2}=\int_0^t{{dt^\prime}\over{\sqrt{1+(\alpha t^\prime/c)^2}}}=(c/\alpha) Sh^{-1}({{\alpha t}\over{c}})$ i.e. $$t=(c/\alpha) Sh(\alpha\tau/c).\eqno{(13)}$$  If this is replaced in the solution for $x(t)$ one obtains $$x=(c^2/\alpha)Ch(\alpha\tau/c). \eqno{(14)}$$

\

It is easy to verify that a Lorentz boost $\pmatrix{x^{0\prime}\cr x^\prime\cr}=\pmatrix{Ch\psi & Sh\psi\cr Sh\psi & Ch\psi\cr}\pmatrix{x^0\cr x}$ shifts the proper time to $\tau^\prime=\tau+{{c}\over{\alpha}}\psi$. 

\

{\it II.4. Rindler coordinates}

\

The usual Rindler coordinates $(\lambda,\xi)$ for the right Rindler wedge (useful for the calculation of the thermal Unruh effect, see section {\bf III}) are defined by $$ct=(c^2/a)e^{a\xi/c^2}Sh(a\lambda/c),\eqno{(15)}$$ $$x=(c^2/a)e^{a\xi/c^2}Ch(a\lambda/c), \eqno{(16)}$$ where $\lambda,\xi\in(-\infty,+\infty)$, $[\xi]=[L]$, $[\lambda]=[T]$, and $a>0$ is a constant with $[a]=[L]/[T]^2$. From this definition, we obtain a hyperbola for each constant $\xi$: $$x^2-c^2t^2=(c^2/a)^2e^{2a\xi/c^2}. \eqno{(17)}$$ Comparing (15) and (16) with (13) and (14), we obtain the relation between the proper acceleration $\alpha$ and $\xi$, and between the proper time $\tau$ and $\lambda$: $$\alpha=ae^{-a\xi/c^2} \eqno{(18)}$$ and $$a\lambda=\alpha\tau. \eqno{(19)}$$ The Jacobian of the transformation (15)-(16) is given by $$J=\pmatrix{x^0_{,\lambda} & x^0_{\xi} \cr x_{,\lambda} & x_{,\xi} \cr}=\pmatrix{ce^{a\xi/c^2}Ch(a\lambda/c) & e^{a\xi/c^2}Sh(a\lambda/c) \cr ce^{a\xi/c^2}Sh(a\lambda/c) & e^{a\xi/c^2}Ch(a\lambda/c) \cr}\eqno{(20)}$$ with $det(J)=ce^{2a\xi/c^2}\in(0,+\infty)$. The inverses of (15)-(16) are: $$\lambda=(c/a)Th^{-1}(ct/x),\eqno{(21)}$$ $$\xi=(c^2/2a)\times ln((a/c^2)^2(x^2-c^2t^2)).\eqno{(22)}$$ Clearly, $R$ with the metric (24) (see below) solves the Einstein equations in vacuum, since $Mink^2$ does.

\

We want to emphasize here that $a$ is {\it not} the proper acceleration: for an arbitrary constant value of $a$ there is a 1-1 correspondence between proper accelerations $\alpha$ and Rindler hyperbole $\xi$. $\lambda$ varies along these lines, so these are the {\it coordinate lines} of $\lambda$. Only for the case $\xi=0$, $a=\alpha$. Since ${{d\alpha}\over{d\xi}}=-(a^2/c^2)e^{-a\xi/c^2}<0$, the larger the value of $a$, more quickly the proper acceleration $\alpha$ decreases to zero as $\xi$ increases beyond $\xi=0$, and more quickly increases to $+\infty$ as $\xi$ decreases below $\xi=0$. For two distinct values of $a$: $a^\prime$ and $a^{\prime\prime}$, the curves $\alpha=\alpha(\xi;a^\prime)$ and $\alpha=\alpha(\xi;a^{\prime\prime})$ intersect at the positive value of $\xi$ given by $\bar{\xi}={{c^2}\over{a^\prime-a^{\prime\prime}}}ln({{a^\prime}\over{a^{\prime\prime}}})$. There is a natural choice to set the scale of accelerations, namely, the {\it Planck acceleration}: $a_{Pl}={{l_{Pl}}\over{t_{Pl}^2}}=({{c^7}\over{\hbar G_N}})^{{{1}\over{2}}}\simeq 5.5\times 10^{51} {{m}\over{sec^2}}$, which, according to eq. (158), would give a {\it universal Unruh temperature}  $T_U={{T_{Pl}}\over{2\pi}}={{\hbar a_{Pl}}\over{2\pi ck_B}}={{c^2}\over{2\pi k_B}}\sqrt{{{\hbar c}\over{G_N}}}\simeq 2.25\times 10^{33} \ {^\circ K}$.

\

Infinite acceleration corresponds to $\xi=-\infty$ (horizon) and zero acceleration corresponds to $\xi=+\infty$ (observer at rest at $x=+\infty$: dashed line in Figure 2). Also, $ct/x=Th(a\lambda/c)$ and so $$ct=Th(a\lambda/c)x\eqno{(23)}$$ i.e. the set of points $\{\lambda=const.\}$ corresponds to a line through the origin in $Mink^2$ space. $\xi$ varies along these lines, so these are the {\it coordinate lines} of $\xi$. Though not in the domain of definition, the limits $\lambda=\pm\infty$ respectively correspond to $ct=\pm x$. $\lambda=0$ is the $t=0$ axis. (See Figure 2.) $(\lambda,\xi)=(0,0)$ is the coordinate origin of the frame, and the line $\xi=0$ its spatial origin. Also, the frame is {\it rigid} in the sense that the proper distance between any two hyperbolae is constant; this can be seen by computing $\Delta x(t)^2$ corresponding to two neighbouring accelerations $\alpha_f$ and $\alpha_b$ to first order in $\delta X=X_f-X_b=c^2(\alpha_f^{-1}-\alpha_b^{-1})=\Delta x(0)$, the result being $\Delta x(t)=\sqrt{1-v_b(t)^2/c^2}\delta X$. Any line $\{\lambda_0=const.\}$ is a Cauchy ``surface'' of $R$ since any inextendible past (future) directed causal curve through any $p\in\{\lambda^\prime=const.>\lambda_0\}$ ($q\in\{\lambda^{\prime\prime}=const.<\lambda_0\}$) intersects $\{\lambda_0=const.\}$. Then $R$ in itself is a {\it globally hyperbolic} spacetime. 

\

A straightforward calculation using $dt=(\partial_\lambda t)d\lambda+(\partial_\xi t)d\xi$ and $dx=(\partial_\lambda x)d\lambda+(\partial_\xi x)d\xi$ leads to the metric $$ds^2=e^{2a\xi/c^2}(c^2d\lambda^2-d\xi^2) \eqno{(24)}$$ which shows that $\lambda$ (or $c\lambda$) is a timelike coordinate while $\xi$ is a spacelike coordinate, and that the Rindler metric is conformal to the Minkowski metric with conformal factor $\Lambda(\lambda,\xi)=e^{a\xi/c^2}$. Along each hyperbola i.e. at each fixed acceleration $\alpha$, the conformal factor remains constant. The metric coefficients are $g_{\lambda\lambda}=-g_{\xi\xi}=e^{2a\xi/c^2}$ and $g_{\lambda\xi}=g_{\xi\lambda}=0$; therefore $\partial_\lambda$ is a Killing vector field with $\vert\vert\partial_\lambda\vert\vert^2=g_{\lambda\lambda}=e^{2a\xi/c^2}$, which is null at the horizon. Also, if $\alpha_1$ and $\alpha_2$ are accelerations corresponding to the values $\xi_1$ and $\xi_2$ of the $\xi$ coordinate, then ${{\alpha_1}\over{\alpha_2}}=e^{-a(\xi_1-\xi_2/c^2)}={{\vert\vert\partial_\lambda\vert_{\xi_2}\vert\vert}\over{\vert\vert\partial_\lambda\vert_{\xi_1}\vert\vert}}$; then $$\alpha(\xi)={{\vert\vert\partial_\lambda\vert_{\xi=0}\vert\vert}\over{\vert\vert\partial_\lambda\vert_{\xi}\vert\vert}}\alpha(0)=ae^{-a\xi/c^2} \eqno{(25)}$$ since $\vert\vert\partial_\lambda\vert_{\xi=0}\vert\vert=1$. So, $\vert\vert\partial_\lambda\vert_{\xi}\vert\vert^{-1}$ is a {\it redshift factor}. (See also subsections {\it II.5.} and {\it II.6.}) At the horizon, if extended, the metric would be singular: $e^{-\infty}=0$.

\

{\it II.5. Dimensionless coordinates}

\

Set $c=1$, choose $a=1$, and define Minkowski and Rindler time and space coordinates without dimensions, respectively $(T,X)$ and $(\eta,\rho)$: $$T=\rho Sh\eta,\eqno{(26)}$$ $$X=\rho Ch\eta,\eqno{(27)}$$ with $T,\eta\in(-\infty,+\infty)$ and $X,\rho\in (0,+\infty)$. (Alternatively, these equations can be obtained from (15) and (16) setting $a\lambda=\eta$, $at=T$, $ax=X$, $c=1$, and $\rho=e^\xi$.) The metric (24) becomes $$ds^2=\rho^2d\eta^2-d\rho^2, \eqno{(28)}$$ which clearly shows that, in $R$, $\eta$ is timelike and $\rho$ is spacelike. 

\

The metric is independent of $\eta$, so $$\partial_\eta=(\partial_\eta T)\partial_T+(\partial_\eta X)\partial_X=X\partial_T+T\partial_X\eqno{(29)}$$ is a {\it Killing vector field} in $R$ and, since $$\vert\vert\partial_\eta\vert\vert ^2=<\partial_\eta,\partial_\eta>=g_{\eta\eta}=\rho^2>0, \eqno{(30)}$$ $\partial_\eta$ is temporal. Therefore, $R$, equipped with the coordinates $\eta$ and $\rho$ (or $\lambda$ and $\xi$) is a {\it static space} since it is stationary: the metric does not depend on the Rindler time $\eta$ (or $\lambda$), and there is no crossed term $d\eta d\rho$ (or $d\lambda d\xi$). From (29), $\partial_\eta$ is the generator of Lorentz boosts in the $X$ direction.

\

It can be easily verified that the hyperbole $\{\rho=const.\}$ are {\it integral curves} or orbits of $\partial_\eta$, that is, if $\vec{T}_p$ is a tangent vector to $\rho=const.$ at $p=(T,X)$, then $\vec{T}_p=\partial_\eta\vert_p.$ In fact, for constant $\rho$, $0={{d\rho^2}\over{dT}}={{d(X^2-T^2)}\over{dT}}=2X{{dX}\over{dT}}-2T$ i.e. ${{dT}\over{dX}}=\vec{T}_{(T,X)}={{X}\over{T}}={{\partial_\eta\vert_T}\over{\partial_\eta\vert_X}}.$

\

So, a hyperbolic motion with given proper acceleration $\rho^{-1}$, asymptotically comes from $\eta=-\infty$ and goes to $\eta=+\infty$. $T=0$ corresponds to $\eta=0$. 

\

Though the coordinates $\eta$ and $\rho$ were derived in the context of the hyperbolic motion of a massive classical point particle, they can be defined independently of any particle motion as curvilinear coordinates in $R$. They are nothing but hyperbolic polar coordinates in $R$. For completeness, we study {\it timelike geodesic motion} in $R$ which is given by the equations $${{d^2y^\mu}\over{d\tau^2}}+\Gamma^\mu_{\nu\rho}{{dy^\nu}\over{d\tau}}{{dy^\rho}\over{d\tau}}=0,\eqno{(31)}$$ with $(y^0,y^1)=(\eta,\rho)$. For the Christoffel symbols $\Gamma^\alpha_{\beta\gamma}={{1}\over{2}}g^{\alpha\delta}(\partial_\beta g_{\gamma\delta}+\partial_\gamma g_{\beta\delta}-\partial_\delta g_{\beta\gamma})$ one obtains: $$\Gamma^\eta_{\eta\eta}=\Gamma^\eta_{\rho\rho}=\Gamma^\rho_{\eta\rho}=\Gamma^\rho_{\rho\rho}=0,$$ $$\Gamma^\eta_{\eta\rho}={{1}\over{\rho}},$$ and $$\Gamma^\rho_{\eta\eta}=\rho.\eqno{(32)}$$ Then the $\eta$- and $\rho$-geodesic equations are, respectively, $$\ddot{\eta}+{{2}\over{\rho}}\dot{\eta}\dot{\rho}=0, \eqno{(33)}$$ and $$\ddot{\rho}+\rho\ddot{\eta}^2=0.\eqno{(34)}$$ The first integral of (33) is $$\rho^2\dot{\eta}=K=const.,\eqno{(35)}$$ with $[K]=[L]^{-1}$, and from the metric $ds^2=d\tau^2$ (see Table I) one obtains $1=\rho^2\dot{\eta}^2-\dot{\rho}^2$, which, from (35), gives $$\dot{\rho}^2={{K^2}\over{\rho^2}}-1.\eqno{(36)}$$ Then, $\dot{\rho}\ddot{\rho}=-K^2\rho^{-3}\dot{\rho}$, which, for $\dot{\rho}\neq 0$, gives the $\rho$-differential equation $$\ddot{\rho}+{{K^2}\over{\rho^3}}=0.\eqno{(37)}$$ (The same equation is obtained from (34) and (35).) Once (37) is integrated, one defines $G(\tau)=2\dot{\rho}/\rho$, and from (33) one obtains the $\eta$-differential equation: $$\ddot{\eta}+G(\tau)\dot{\eta}=0. \eqno{(38)}$$ The solution of the autonomous equation (37) is $$\rho(\tau)=\sqrt{K^2(\beta+\tau)^2-1} \eqno{(39)}$$ where $\beta$ is an integration constant with $[\beta]=[L]$. $\rho$ is real for $$\vert K(\beta+\tau)\vert\geq 1 \eqno{(40)}$$ and {\it reaches the horizon} $\rho=0$ for $$\tau=\bar{\tau}=\pm\vert K\vert^{-1}-\beta. \eqno{(41)}$$ For $G(\tau)$ one obtains $$G(\tau)={{2K^2(\beta+\tau)}\over{\rho^2}},$$ and defining $u=\dot{\eta}$, which is positive for a future directed geodesic, one has the linear equation $\dot{u}+G(\tau)u=0$ which is trivially integrated. The solution for $\eta$ is $$\eta(\tau)=u(\tau_0)(K^2(\beta+\tau_0)^2-1) \ ln\vert{{(K(\beta+\tau)+1)}\over{(K(\beta+\tau)-1)}}{{(K(\beta+\tau_0)-1)}\over{(K(\beta+\tau_0)+1)}}\vert \ .\eqno{(42)}$$ Then, for $\tau=\bar{\tau}=K^{-1}-\beta$, $\eta$ has a logarithmic divergence: $$\eta(K^{-1}-\beta)=b+a\times(+\infty)=+\infty \eqno{(43)}$$ for both $K>0$ and $K<0$, with $a=u(\tau_0)(K^2(\beta+\tau_0)^2-1)>0$ and $b=a \ ln\vert{{K(\beta+\tau_0)-1}\over{K(\beta+\tau_0)+1}}\vert \ .$ This fact, together with $\rho(\bar{\tau})=0$, prove the {\it geodesic incompleteness} of the Rindler space.

\

Notice that: 

\

i) The proper time of the total geodesic motion from the horizon with $\eta=-\infty$ to the horizon with $\eta=+\infty$ is finite: $$\Delta\tau=(\vert K\vert^{-1}-1)-(-\vert K\vert^{-1}-1)=2/\vert K\vert.\eqno{(44)}$$ 

\

ii) The curvature tensor (and obviously all its invariants) vanishes: $$R_{\eta\rho\eta\rho}=g_{\eta\eta}R^\eta_{\rho\eta\rho}=\rho^2(-\partial_\rho\Gamma^\eta_{\rho\eta}-\Gamma^\eta_{\rho\eta}\Gamma^\eta_{\eta\rho})=\rho^2(-\partial_\rho(1/\rho)-(1/\rho)^2)=0 \eqno{(45)}$$ (this is so because the gravity ``seen'' by the Rindler observer is due to its acceleration with respect to the inertial frame $Mink^2$).

\

Then the singularity at $(\eta,\rho)=(\pm\infty,0)$ is a coordinate singularity and therefore $R$ is {\it regular} there and can be extended: precisely $Mink^2$ is its {\it maximal analytic extension}.

\

{\it II.6.} $Mink^2$ {\it spacetime as the maximal analytic extension of R}

\

(Obviously the result is also valid in the 4-dimensional case, i.e. for $Mink^4$, adding the variables $x^2$ and $x^3$.)

\

The process consists of a successive change of coordinates given by {\it analytic functions} of their arguments in their domains of definition, starting from $R$ with coordinates $\eta$ and $\rho$, and metric $ds^2=\rho^2d\eta^2-d\rho^2$, where $det(g_{\alpha\beta}(\eta,\rho))=-\rho^2$. 

\

i) Define light cone or null coordinates $$\tilde{u}=\tilde{u}(\eta,\rho):=\eta-ln\rho, \ \tilde{v}=\tilde{v}(\eta,\rho):=\eta+ln\rho, \ [\tilde{u}]=[\tilde{v}]=[L]^0, \ \tilde{u},\tilde{v}\in(-\infty,+\infty).\eqno{(46)}$$ The $\tilde{u}$ ,$\tilde{v}$ coordinates of the future ($(\rho,\eta)_+=(0,+\infty)$) and past ($(\rho,\eta)_-=(0,-\infty)$) horizons are respectively $(\tilde{u},\tilde{v})_+=(+\infty,\tilde{v}<\infty)$ and $(\tilde{u},\tilde{v})_-=(\tilde{u}<\infty,\tilde{v}=-\infty)$. (These coordinates are the dimensionless version of the coordinates $\bar{v}=c\lambda+\xi$ and $\bar{u}=c\lambda-\xi$ to be defined in section 1.8, with $a=1$, $\eta=\lambda$, and $\rho=e^\xi$.) The inverse transformation is $$\eta={{\tilde{v}+\tilde{u}}\over{2}}, \ \rho=e^{{{\tilde{v}-\tilde{u}}\over{2}}}. \eqno{(47)}$$ Then $d\eta={{1}\over{2}}(d\tilde{v}+d\tilde{u})$, $d\rho={{1}\over{2}}\rho(d\tilde{v}-d\tilde{u})$ and therefore $$ds^2=\rho^2d\tilde{v}d\tilde{u}=e^{\tilde{v}-\tilde{u}}d\tilde{v}d\tilde{u},\eqno{(48)}$$ with metric tensor $$g_{\alpha\beta}(\tilde{u},\tilde{v}))={{1}\over{2}}\pmatrix{0 & e^{\tilde{v}-\tilde{u}} \cr e^{\tilde{v}-\tilde{u}} & 0 \cr}, \ det(g_{\alpha,\beta}(\tilde{u},\tilde{v}))=-{{1}\over{4}}e^{2(\tilde{v}-\tilde{u})}. \eqno{(49)}$$ At the horizons the metric does not exist since $$g_{\alpha,\beta}(\tilde{u},\tilde{v})_\pm=0.\eqno{(50)}$$ 

\

ii) Now define $$V=V(\tilde{u},\tilde{v}):=e^{\tilde{v}}, \ U=U(\tilde{u},\tilde{v}):=e^{-\tilde{u}}, \ [V]=[U]=[L]^0, \ V,U\in(0,+\infty),\eqno{(51)},$$ with inverse $$\tilde{v}=lnV, \ \tilde{u}=-lnU \eqno{(52)}$$ and metric $$ds^2=-dVdU, \eqno{(53)}$$ i.e. with metric tensor $$g_{\alpha\beta}(U,V)=\pmatrix{0 & -{{1}\over{2}} \cr -{{1}\over{2}} & 0 \cr}, \ det(g_{\alpha\beta}(U,V))=-{{1}\over{4}}.\eqno{(54)}$$ The $U$, $V$ coordinates of the horizons are now $(U,V)_+=(0,e^{\tilde{v}}<\infty)$ and $(U,V)_-=(e^{-\tilde{u}}<\infty,0)$: the metric is {\it regular} and can be {\it extended} beyond them.

\

iii) Defining $$X^0=X^0(U,V):={{1}\over{2}}(V-U), \ X^1=X^1(U,V):={{1}\over{2}}(V+U), \ [X^0]=[X^1]=[L]^0, \ X^0,X^1\in(-\infty,+\infty)\eqno{(55)}$$ one obtains the metric $$ds^2=(dX^0)^2-(dX^1)^2\eqno{(56)}$$ i.e. the metric tensor $$g_{\alpha\beta}(X^0,X^1)=\pmatrix{1 &0 \cr 0 & -1 \cr}, \ det(g_{\alpha,\beta}(X^0,X^1))=-1.\eqno{(57)}$$ That is, $Mink^2$ spacetime. (The whole process is summarized in Figure 3.)

\

\centerline{\epsfxsize=65ex\epsfbox{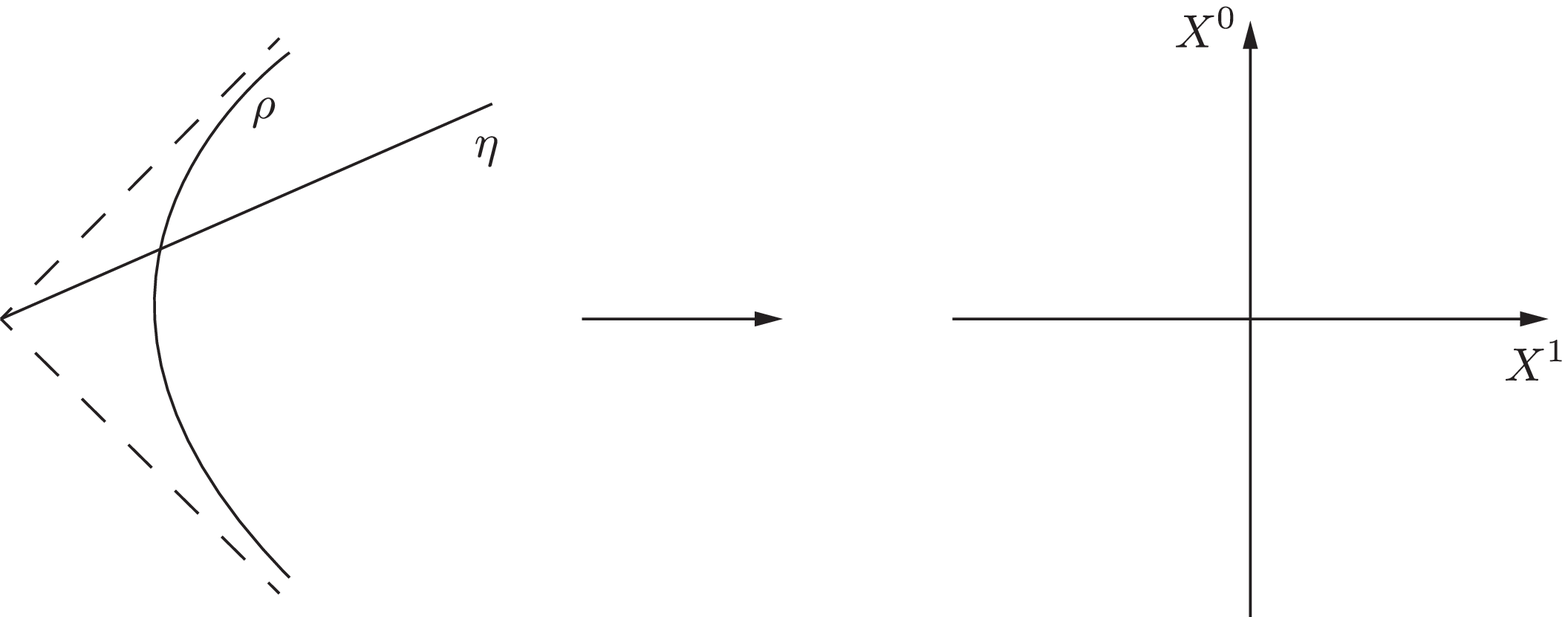}}

\

\centerline{Figure 3. Minkowski space as the maximal analytic extension of the right Rindler wedge}

\

{\it II.7. Left, future, and past wedges}

\

Formally, there exist trajectories, integral curves of the vector field ${{\partial}\over{\partial \eta}}$, in the {\it left Rindler wedge} $L$ ($X<\pm T$), {\it future Rindler wedge} $F$ ($T>\pm X$), and {\it past Rindler wedge} $P$ ($T<\pm X$) with coordinates $\rho\in (0,+\infty)$ and $\eta\in (-\infty,+\infty)$ as in $R$. Each trajectory ``moves" in Rindler time $\eta$ and along a fixed $\rho$ from $\eta=-\infty$ towards $\eta=+\infty$. In $L$, which by the same reasons as in {\it II.4} and {\it II.5} for $R$ is also a globally hyperbolic static spacetime, trajectories are timelike and can be understood as antiparticles moving backwards in Minkowski time with negative proper acceleration. In $F$ and $P$ trajectories are spacelike and do not correspond to particle motions. Also, in $F$ and $P$ the $T$-axis corresponds to $\eta=0$. The relation with the Minkowski coordinates, tangent vectors, and metrics is given in Table I, and typical trajectories are shown in Figure 4. In Figure 5, $\eta=-\infty$ and $T\geq X$ are respectively the horizon and invisible region for the observer (antiparticle) with the indicated trajectory in $L$. 

\

\centerline{\epsfxsize=45ex\epsfbox{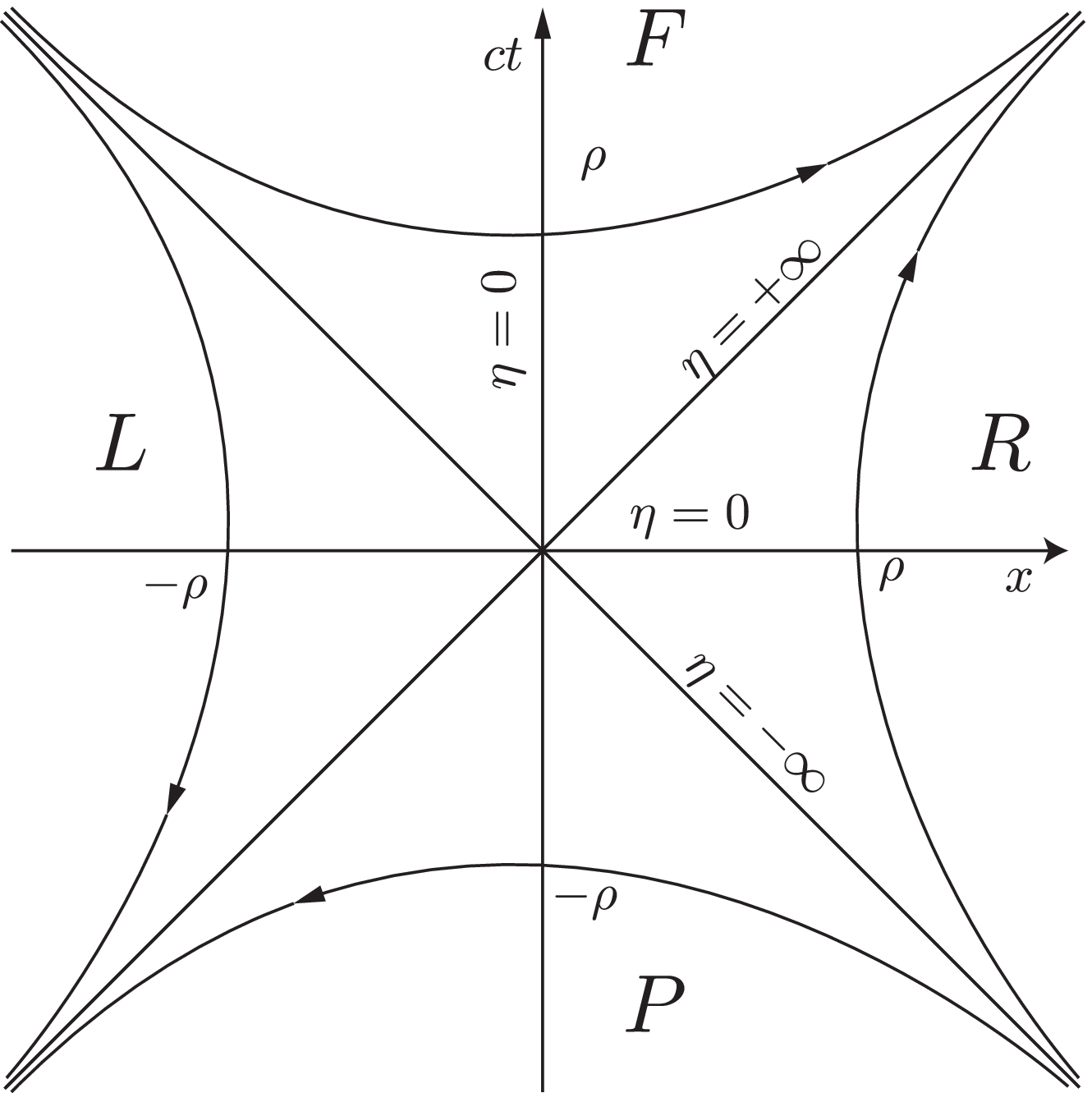}} 

\

\centerline{Figure 4. Trajectories in the four Rindler wedges}  

\

\centerline{\epsfxsize=45ex\epsfbox{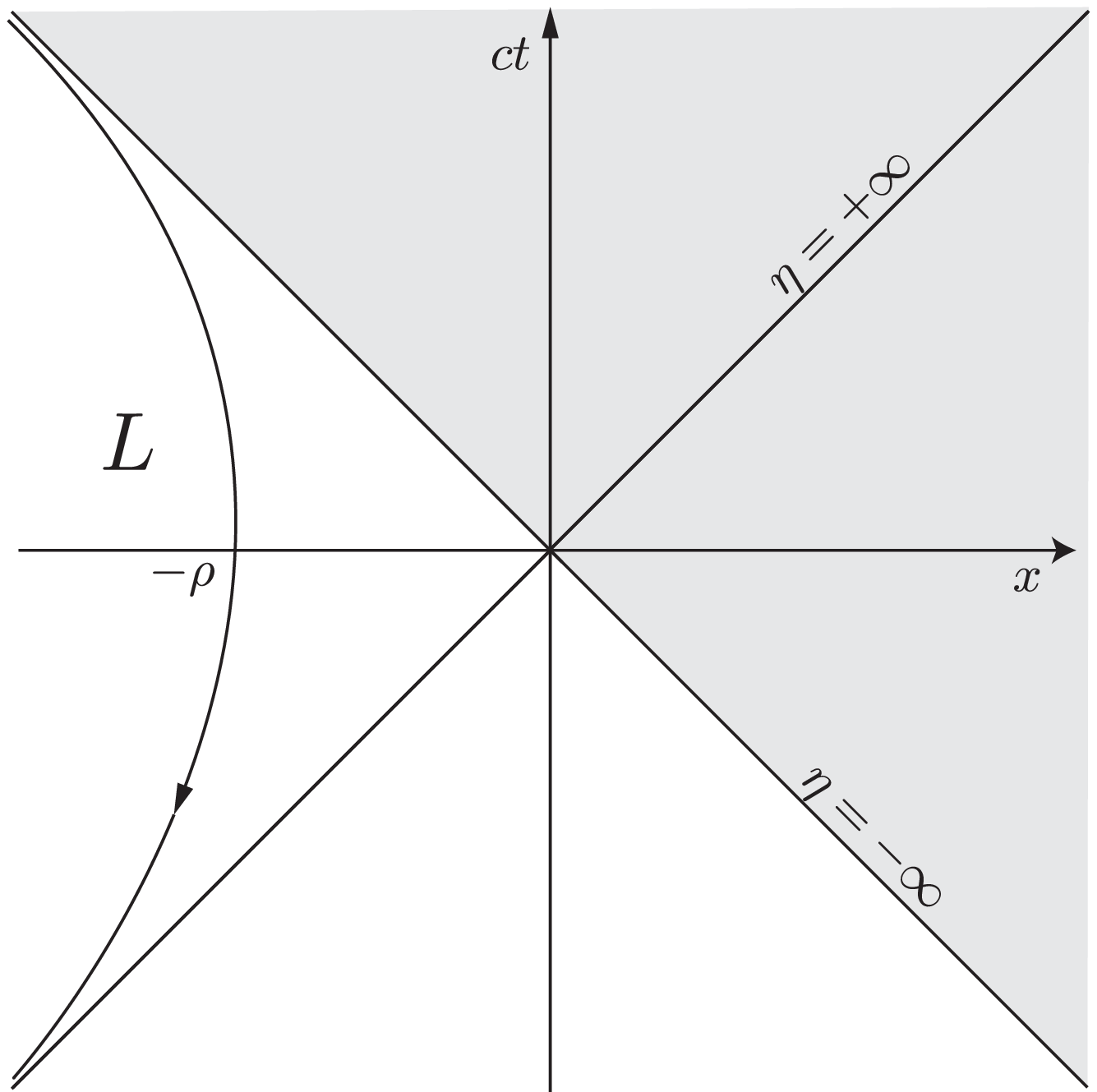}} 

\

\centerline{Figure 5. Horizon and invisible region for a trajectory in the left Rindler wedge}

\

Inspection of the metric in the last column of Table I, shows that:

\

i) $\eta$ and $\rho$ are respectively timelike and spacelike coordinates in $R$ and $L$, while in $F$ and $P$ $\eta$ is spacelike and $\rho$ is timelike. So, at the horizons, a timelike (spacelike) coordinate changes into a spacelike (timelike) coordinate, as it happens at the horizon of a Schwarzschild black hole. 

\

ii) ${{\partial}\over{\partial \eta}}$ is a Killing vector field in the four wedges, timelike in $R$ and $L$ but spacelike in $F$ and $P$. On the horizon, $\partial_\eta$ is null i.e. $\vert\vert\partial_\eta\vert\vert^2=0$, since $\rho^{-1}\vert_{hor.}=\alpha\vert_{hor.}=+\infty$, and is orthogonal to it, since $<\partial_\eta,(T,X)>=<(X,T),(T,X)>=\eta_{00}XT+\eta_{11}TX=XT-TX=0$. This is the reason why $T=\pm X$ are called Killing horizons.

\

iii) ${{\partial}\over{\partial \rho}}$ is timelike in $F$ and $P$ and spacelike in $R$ and $L$, but it is not a Killing field since ${{\partial}\over{\partial \rho}}g_{\eta\eta}\vert_{F,P}=-2\rho\neq 0$. So, $F$ and $P$ are not stationary spaces and therefore also non static. They are respectively called {\it expanding} and {\it contracting degenerate Kasner spaces}.

\

iv) In terms of Minkowski coordinates, trajectories in $R$ and $L$ are interchanged by a $\hat{C}\hat{P}\hat{T}$ transformation: $\hat{P}:X\to -X$, $\hat{T}:T\to -T$, and $\hat{C}$: a particle moving forward in $T$-time goes to an antiparticle moving backwards in $T$-time. 

\

v) In terms of Rindler coordinates, trajectories in $R$ and $L$ are interchanged only by a $\hat{C}\hat{P}$ transformation, since $\eta$ is the same, $\hat{P}:\rho\to -\rho$, and $\hat{C}$: particle goes to antiparticle but now moving forward in $\eta$-time.

$$\matrix{& X & T & X^2-T^2 & {{T}\over{X}} & \partial_\eta & \vert\vert\partial_\eta\vert\vert^2 & \partial_\rho & \vert\vert\partial_\rho\vert\vert^2 & ds^2 \cr R & \rho Ch\eta & \rho Sh\eta & \rho^2 & Th\eta & X\partial_T+T\partial_X & \rho^2 & {{X\partial_X+T\partial_T}\over{\sqrt{X^2-T^2}}} & -1 & \rho^2d\eta^2-d\rho^2\cr L & -\rho Ch\eta & -\rho Sh\eta & \rho^2 & Th\eta & X\partial_T+T\partial_X & \rho^2 & -{{X\partial_X+T\partial_T}\over{\sqrt{X^2-T^2}}} & -1 & \rho^2d\eta^2-d\rho^2\cr F & \rho Sh\eta & \rho Ch\eta & -\rho^2 & Coth\eta & X\partial_T+T\partial_X & -\rho^2 & {{X\partial_X+T\partial_T}\over{\sqrt{T^2-X^2}}} & +1 & d\rho^2-\rho^2d\eta^2\cr P & -\rho Sh\eta & -\rho Ch
\eta & -\rho^2 & Coth\eta & X\partial_T+T\partial_X & -\rho^2 & -{{X\partial_X+T\partial_T}\over{\sqrt{T^2-X^2}}}& +1 & d\rho^2-\rho^2d\eta^2\cr}$$ $$Table \ I$$

\

{\it II.8. Light cone coordinates}

\

From the light cone coordinates $u$ and $v$ in Minkowski space ((eq. 3)), defining $$\bar{u}=c\lambda-\xi, \ \ \ \bar{v}=c\lambda + \xi \eqno{(58)}$$ ([$\bar{u}$]=[$\bar{v}$]=[$L$]) with inverses $$\lambda={{\bar{v}+\bar{u}}\over{2c}}, \ \ \ \xi={{\bar{v}-\bar{u}}\over{2}},\eqno{(59)}$$ we obtain for $u=u(\bar{u})$, $v=v(\bar{v})$, and $ds^2$ the expressions in Table II. In particular: $$u=0 \Longleftrightarrow \bar{u}=+\infty, \ \ \ v=0 \Longleftrightarrow \bar{v}=-\infty, \eqno{(60)}$$ $$u=const.\Longleftrightarrow \bar{u}=const., \ \ \ v=const.\Longleftrightarrow\bar{v}=const. \eqno{(61)}$$ For the proper acceleration one obtains $$\alpha=ae^{-{{a}\over{2c^2}}(\bar{v}-\bar{u})}. \eqno{(62)}$$ $$\matrix{& R & L & F & P \cr t& ca^{-1}e^{a\xi/c^2}Sh(a\lambda/c) & -ca^{-1}e^{a\xi/c^2}Sh(a\lambda/c) & ca^{-1}e^{a\xi/c^2}Ch(a\lambda/c) & -ca^{-1}e^{a\xi/c^2}Ch(a\lambda/c)\cr x & c^2a^{-1}e^{a\xi/c^2}Ch(a\lambda/c) & -c^2a^{-1}e^{a\xi/c^2}Ch(a\lambda/c) & c^2a^{-1}e^{a\xi/c^2}Sh(a\lambda/c) & -c^2a^{-1}e^{a\xi/c^2}Sh(a\lambda/c) \cr u & -c^2a^{-1}e^{-a\bar{u}/c^2} & c^2a^{-1}e^{-a\bar{u}/c^2} & c^2a^{-1}e^{-a\bar{u}/c^2} & -c^2a^{-1}e^{-a\bar{u}/c^2}\cr v & c^2a^{-1}e^{a\bar{v}/c^2} & -c^2a^{-1}e^{a\bar{v}/c^2} &  c^2a^{-1}e^{a\bar{v}/c^2} & -c^2a^{-1}e^{a\bar{v}/c^2} \cr ds^2 & e^{a(\bar{v}-\bar{u})/c^2}d\bar{v}d\bar{u} & e^{a(\bar{v}-\bar{u})/c^2}d\bar{v}d\bar{u} & -e^{a(\bar{v}-\bar{u})/c^2}d\bar{v}d\bar{u} & -e^{a(\bar{v}-\bar{u})/c^2}d\bar{v}d\bar{u}\cr}$$ $$Table \ II$$

\

$\bar{u}$ and $\bar{v}$, both taking values in $(-\infty,+\infty)$, are the light cone coordinates in Rindler space. In Figure 6 we plot some lines of constant $\bar{u}$'s and $\bar{v}$'s. 

\centerline{\epsfxsize=55ex\epsfbox{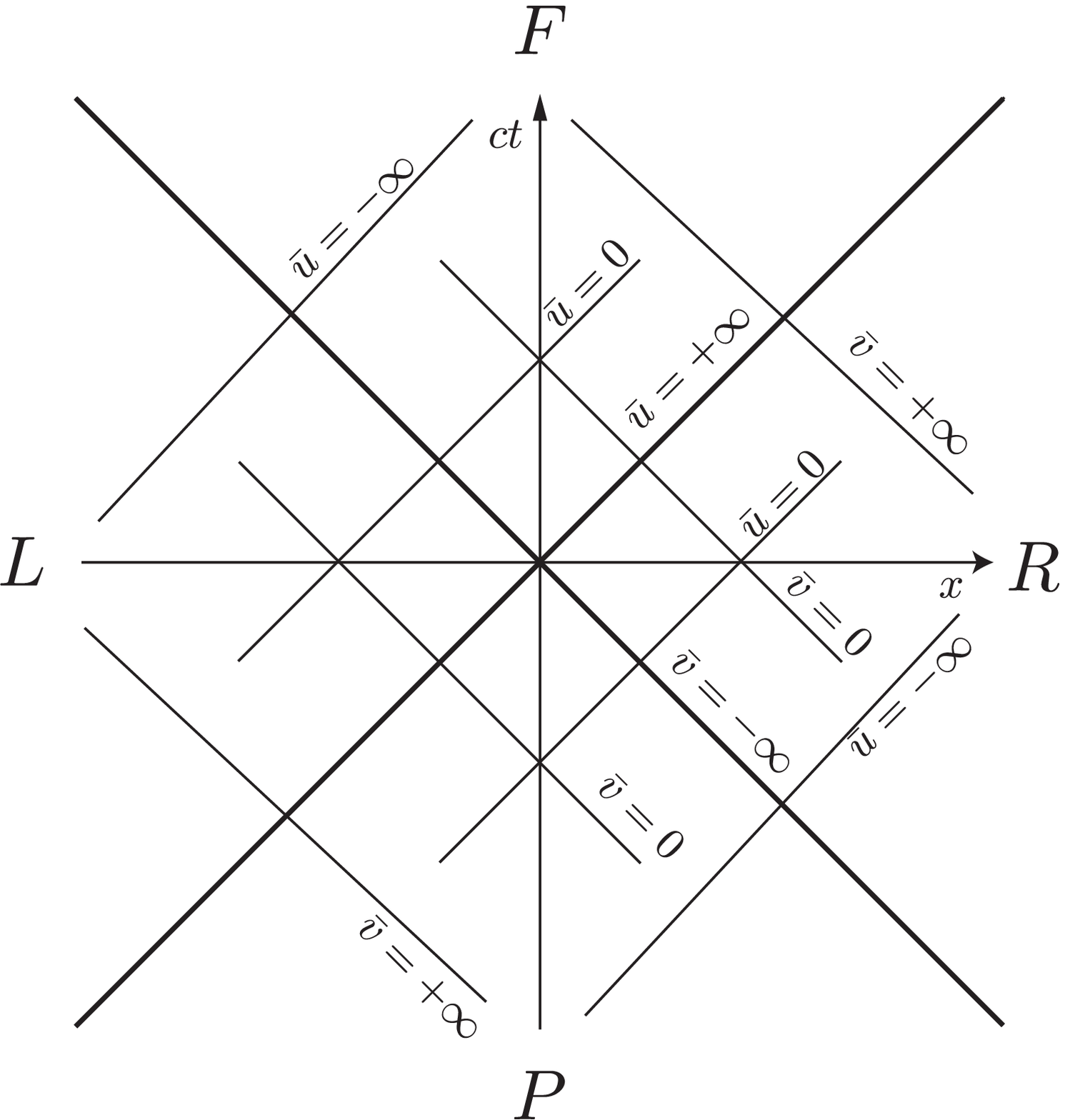}} 

\

\centerline{Figure 6. Rindler light cone coordinate lines}

\

For later use, we give the expressions for the positive frequency right ($R$) and left ($L$) right ($\rightarrow$ or $k>0$) and left ($\leftarrow$ or $k<0$) moving unnormalized solutions of the Klein-Gordon equation $${{\partial^2\phi}\over{\partial\bar{u}\partial\bar{v}}}=0\eqno{(63)}$$ for a free massless real scalar field living in the two dimensional Minkowski (and therefore in the four Rindler wedges) space: $$\phi^R_k=e^{i(k\xi-\omega\lambda)}=e^{-i\omega(\lambda-sg(k)\xi/c)}=\pmatrix{e^{-i\omega(\lambda-\xi/c)}=e^{-i\omega\bar{u}/c}=\phi^R_{\rightarrow}\cr e^{-i\omega(\lambda+\xi/c)}=e^{-i\omega\bar{v}/c}=\phi^R_{\leftarrow}},\eqno{(64)}$$ $$\phi^L_k=e^{i(k\xi+\omega\lambda)}=e^{-i\omega(\lambda+sg(k)\xi/c)}=\pmatrix{e^{i\omega(\lambda+\xi/c)}=e^{i\omega\bar{v}/c}=\phi^L_{\rightarrow}\cr e^{i\omega(\lambda-\xi/c)}=e^{i\omega\bar{u}/c}=\phi^L_{\leftarrow}},\eqno{(65)}$$ where $\omega=c\vert k\vert>0$, $k\in(-\infty,+\infty)$, $\phi_k^R=0$ in $L$ and $\phi_k^L=0$ in $R$, and $$i\partial_\lambda\phi^R_{\rightarrow}=\omega\phi^R_{\rightarrow}, \  \ i\partial_\lambda\phi^R_{\leftarrow}=\omega\phi^R_{\leftarrow}, \  \ i\partial_{(-\lambda)}\phi^L_{\rightarrow}=\omega\phi^L_{\rightarrow}, \  \ i\partial_{(-\lambda)}\phi^L_{\leftarrow}=\omega\phi^L_{\leftarrow}.\eqno{(66)}$$ So, in $R$, right (left) moving solutions move along constant values of $\bar{u}$ ($\bar{v}$), and the opposite occurs in $L$: right (left) moving solutions move along constant values of $\bar{v}$ ($\bar{u}$). The wave equation (63) is a consequence of the invariance of the 2-dimensional D'Alambertian $g^{\mu\nu}D_\mu D_\nu$, which leads to 
$$(c^{-2}\partial^2_t-\partial^2_x)\phi=(c^{-2}\partial^2_\lambda-\partial^2_\xi)\phi=\partial_{\bar{u}}\partial_{\bar{v}}\phi=0.\eqno{(67)}$$ Together, the sets $\{\phi^R_k\}$ and $\{\phi^L_k\}$ (taking only their left moving or right moving, or right (left) for $R$ and left (right) for $L$, parts) are a complete set of solutions in the whole Minkowski spacetime, including the wedges $F$ and $P$. This is because any line of constant $\lambda$ through $R$ {\it and} $L$ is a Cauchy ``surface'' for the whole $Mink^2$.

\

We can rewrite $\phi^R_\rightarrow(\lambda,\xi)$, $\phi^L_\leftarrow(\lambda,\xi)^*$, $\phi^R_\leftarrow(\lambda,\xi)^*$, and $\phi^L_\rightarrow(\lambda,\xi)$ as follows: $$\phi^R_\rightarrow(\lambda,\xi)=e^{-i\omega\bar{u}/c}=(e^{-a\bar{u}/c^2})^{i\omega c/a}=(-au/c^2)^{i\omega c/a}=(a/c^2)^{i\omega c/a}(-u)^{i\omega c/a}=(a/c^2)^{i\omega c/a}(-ct+x)^{i\omega c/a}, \eqno{(68)}$$ $$\phi^L_\leftarrow(\lambda,\xi)^*=e^{-i\omega\bar{u}/c}=(e^{-a\bar{u}/c^2})^{i\omega c/a}=(au/c^2)^{i\omega c/a}=(a/c^2)^{i\omega c/a}(-1)^{i\omega c/a}(-u)^{i\omega c/a}$$ $$=(a/c^2)^{i\omega c/a}(-1)^{i\omega c/a}(-ct+x)^{i\omega c/a}, \eqno{(69)}$$ $$\phi^R_\leftarrow(\lambda,\xi)^*=e^{i\omega\bar{v}/c}=(e^{a\bar{v}/c^2})^{i\omega c/a}=(av/c^2)^{i\omega c/a}=(a/c^2)^{i\omega c/a}(v)^{i\omega c/a}=(a/c^2)^{i\omega c/a}(ct+x)^{i\omega c/a}, \eqno{(70)}$$ $$\phi^L_\rightarrow(\lambda,\xi)=e^{i\omega\bar{v}/c}=(e^{a\bar{v}/c^2})^{i\omega c/a}=(-av/c^2)^{i\omega c/a}=(a/c^2)^{i\omega c/a}(-1)^{i\omega c/a}(v)^{i\omega c/a}$$ $$=(a/c^2)^{i\omega c/a}(-1)^{i\omega c/a}(ct+x)^{i\omega c/a}. \eqno{(71)}$$ In (69) and (71) (respectively negative and positive frequency left modes) there is an apparent ambiguity for the choice in $(-1)^{i\omega c/a}$: $-1=e^{i\pi}$ or $-1=e^{-i\pi}$; however, the fact that $v>0$ ($<0$) and $u<0$ ($>0$) in $R$ ($L$), and that to keep positive frequencies in the Minkowski linear combinations (74) and (75) (see below) one needs $Im \ u$, $Im \ v$ $<0$, the analytic continuation from $R$ to $L$ (for which we write $u=Re \ u+iIm \ u$ and $v=Re \ v+iIm \ v$) requires $Im \ u$ $<0$ and therefore $-1=e^{-i\pi}$ in (69) (the continuation must be done in the lower half of the complex $u$-plane), and $Im \ (-v)$ $>0$ and therefore $-1=e^{i\pi}$ in (71) (the continuation must be done in the upper half of the complex $v$-plane). So, we obtain: $$\phi^L_\leftarrow(t,x)^*=(a/c^2)^{i\omega c/a}(e^{-i\pi})^{i\omega c/a}(-ct+x)^{i\omega c/a}=e^{\pi\omega c/a}(a/c^2)^{i\omega c/a}(-ct+x)^{i\omega c/a} \eqno{(72)}$$ and $$\phi^L_\rightarrow(t,x)=(a/c^2)^{i\omega c/a}(e^{i\pi})^{i\omega c/a}(ct+x)^{i\omega c/a}=e^{-\pi\omega c/a}(a/c^2)^{i\omega c/a}(ct+x)^{i\omega c/a}. \eqno{(73)}$$ Therefore, normalized positive frequency solutions in the whole Minkowski space but written in terms of $L$ and $R$ positive and negative frequency solutions are given by the following expressions: $$\varphi^{(1)}_k(t,x)/N=e^{\pi\omega c/2a}\phi^R_k(t,x)+e^{-\pi\omega c/2a}\phi^L_{-k}(t,x)^*=2e^{\pi\omega c/2a}(a/c^2)^{i\omega c/a}(-ct+x)^{i\omega c/a}, \eqno{(74)}$$ $$\varphi^{(2)}_k(t,x)/N=e^{-\pi\omega c/2a}\phi^R_{-k}(t,x)^*+e^{\pi\omega c/2a}\phi^L_k(t,x)=2e^{-\pi\omega c/2a}(a/c^2)^{i\omega c/a}(ct+x)^{i\omega c/a}.\eqno{(75)}$$ To simplify the calculation of the normalization factor $N$, momentarily we assume discrete values for $k$ with scalar products $(\varphi^{(a)}_k,\varphi^{(a)}_{k^\prime})=\delta_{kk^\prime}$, $a=1,2$; $(\phi^{L,R},\phi^{L,R})=-((\phi^{L,R})^*,(\phi^{L,R})^*)=1$, and $(\phi^{L,R},(\phi^{L,R})^*)=0$. Neglecting an irrelevant phase we obtain $N=1/\sqrt{2Sh({{\pi\omega c}\over{a}})}$ and so $$\varphi^{(1)}_k(t,x)={{1}\over{\sqrt{2Sh({{\pi\omega c}\over{a}})}}}(e^{\pi\omega c/2a}\phi^R_k(t,x)+e^{-\pi\omega c/2a}\phi^L_{-k}(t,x)^*)=\sqrt{{{2}\over{Sh({{\pi\omega c}\over{a}})}}}e^{\pi\omega c/2a}(a/c^2)^{i\omega c/a}(-ct+x)^{i\omega c/a}, \eqno{(76)}$$ and $$\varphi^{(2)}_k(t,x)={{1}\over{\sqrt{2Sh({{\pi\omega c}\over{a}})}}}(e^{-\pi\omega c/2a}\phi^R_{-k}(t,x)^*+e^{\pi\omega c/2a}\phi^L_k(t,x))=\sqrt{{{2}\over{Sh({{\pi\omega c}\over{a}})}}}e^{-\pi\omega c/2a}(a/c^2)^{i\omega c/a}(ct+x)^{i\omega c/a}.\eqno{(77)}$$ These expressions will be used in the derivation of the Unruh effect in section {\bf III}, following the original strategy.

\

{\it II. 9. Penrose spaces (diagrams) of Minkowski and Rindler spaces}

\

{\it II.9.1. Minkowski space}

\

The Penrose diagrams or Penrose spaces capture the global properties and causal structures of a given spacetime. For the 4-dimensional Minkowski space, it is constructed through the following six steps:

\

$$cartesian \ \ coordinates\buildrel{(i)}\over\longrightarrow spherical \ \ coordinates\buildrel{(ii)}\over\longrightarrow lightcone \ \ coordinates\buildrel{(iii)}\over\longrightarrow finite \ \ lightcone$$ $$coordinates\buildrel{(iv)}\over\longrightarrow cartesian \ \ coordinates\buildrel{(v)}\over\longrightarrow conformal \ \ metric\buildrel{(vi)}\over\longrightarrow compactified \ \ space \eqno{(78)}$$ In detail:

\

(i) $x^\mu$ in eq. (1)$\to (t,r,\theta,\varphi)$, with $r\geq 0$, $\theta\in [0,\pi]$, and $\varphi\in [0,2\pi)$. The $z$-axis ($\theta=0,\pi$, including $r=0$) is a coordinate singularity, which would require a second chart; however, we'll proceed with it. The metric becomes $$ds^2=c^2dt^2-dr^2-r^2d^2\Omega, \ \ d^2\Omega=d\theta^2+sin^2\theta d\varphi^2$$ i.e. $$g_{\mu\nu}\vert_{t,r,\theta,\varphi}=diag(1,-1,-r^2,-r^2sin^2\theta). \eqno{(79)}$$ We have an open an unbounded space.

\

(ii) $(t,r,\theta,\varphi)\to (u,v,\theta,\varphi)$ with $$u=ct-r, \ \ v=ct+r; \ \ u,v\in(-\infty,+\infty),$$ and inverses $$ct={{v+u}\over{2}}, \ \ r={{v-u}\over{2}}, \ \ v\geq u.\eqno{(80)}$$ We have: $$u=const.\Longrightarrow ct=r+const.: \ \ outgoing \ \ light \ \ rays, \ \ u\to -\infty : \ \ past \ \ null \ \ infinity;$$ $$v=const.^\prime \Longrightarrow ct=-r+const.^\prime: \ \ ingoing \ \ light \ \ rays, \ \ v\to +\infty : \ \ future \ \ null \ \ infinity.\eqno{(81)}$$ The metric is $$ds^2=dvdu-{{(v-u)^2}\over{4}}d^2\Omega$$ i.e. $$g_{\mu\nu}\vert_{(u,v,\theta,\varphi)}=\pmatrix{0 & {{1}\over{2}} & 0 & 0 \cr {{1}\over{2}} & 0 & -{{(v-u)^2}\over{4}} & 0 \cr 0 & 0 & 0 & -{{(v-u)^2}\over{4}}sin^2\theta\cr}.\eqno{(82)}$$ As in (i), we have still an open and unbounded space.

\

(iii) $(u,v,\theta,\varphi)\to (u^\prime,v^\prime,\theta,\varphi)$ with $$u^\prime=2tg^{-1}u/L, \ \ v^\prime=2tg^{-1}v/L, \ \ u^\prime,v^\prime\in (-\pi,\pi), \ \ v^\prime\geq u^\prime,$$ and inverses $${{u}\over{L}}=tg{{u^\prime}\over{2}}, \ \ {{v}\over{L}}=tg{{v^\prime}\over{2}}.\eqno{(83)}$$ $L$ is an arbitrary length scale (see below).

\

We have: $$u^\prime\to -\pi: \ \ past \ \ null \ \ infinity,$$ $$v^\prime\to\pi: \ \ future \ \ null \ \ infinity. \eqno{(84)}$$ The metric becomes $$ds^2={{L^2}\over{4cos^2({{u^\prime}\over{2}})cos^2({{v^\prime}\over{2}})}}(du^\prime dv^\prime-sin^2({{v^\prime-u^\prime}\over{2}})d^2\Omega),$$ i.e. $$g_{\mu\nu}\vert_{(u^\prime,v^\prime,\theta,\varphi)}={{L^2}\over{4cos^2({{u^\prime}\over{2}})cos^2({{v^\prime}\over{2}})}}\pmatrix{0 & {{1}\over{2}} & 0 & 0 \cr {{1}\over{2}} & 0 & 0 & 0 \cr 0 & 0 & -sin^2({{v^\prime-u^\prime}\over{2}}) & 0 \cr 0 & 0 & 0 & -sin^2({{v^\prime-u^\prime}\over{2}})sin^2\theta\cr}.\eqno{(85)}$$ We have now an open but bounded space; see Figure 7. In it, each point, except at $r=0$, is a 2-sphere with radius ${{L}\over{2}}(tg({{v^\prime}\over{2}})-tg({{u^\prime}\over{2}}))$. 

\

\centerline{\epsfxsize=55ex\epsfbox{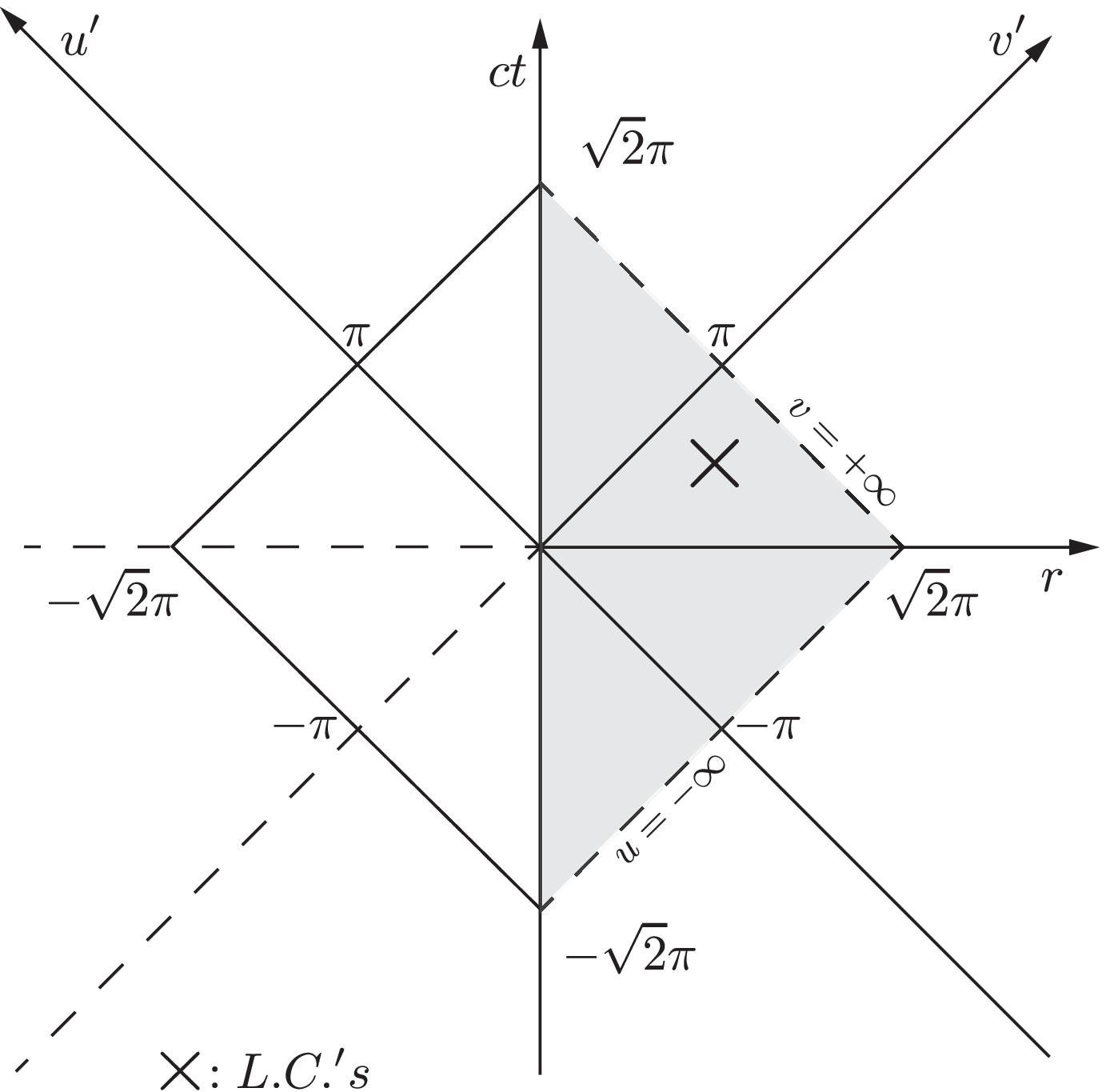}} 

\

\centerline{Figure 7. Open bounded Minkowski space with light cone coordinates}

\

(iv) $(u^\prime,v^\prime,\theta,\varphi)\to (cT,R,\theta,\varphi)$ with $$cT=L{{v^\prime+u^\prime}\over{2}}, \ \ R=L{{v^\prime-u^\prime}\over{2}}, \ \ cT\in(-\pi,\pi)\times L, \ \ R\in[0,\pi)\times L,$$ and inverses $$u^\prime={{cT-R}\over{L}}, \ \ v^\prime={{cT+R}\over{L}}.\eqno{(86)}$$ We have: $${{cT}\over{L}}={{R}\over{L}}-\pi: \ \ past \ \ null \ \ infinity,$$ $${{cT}\over{L}}=-{{R}\over{L}}+\pi: \ \ future \ \ null \ \ infinity.\eqno{(87)}$$ The metric is $$ds^2={{1}\over{(cos({{cT}\over{L}})+cos({{R}\over{L}}))^2}}(c^2dT^2-(dR^2+L^2sin^2({{R}\over{L}})d^2\Omega))$$ i.e. $$g_{\mu\nu}\vert_{(T,R,\theta,\varphi)}={{1}\over{(cos({{cT}\over{L}})+cos({{R}\over{L}}))^2}}\pmatrix{1 & 0 & 0 & 0 \cr 0 & -1 & 0 & 0 & \cr 0 & 0 & -L^2sin^2({{R}\over{L}}) & 0 \cr 0 & 0 & 0 & -L^2sin^2({{R}\over{L}})sin^2\theta\cr}.\eqno{(88)}$$ The quantity $$dl^2=dR^2+L^2sin^2({{R}\over{L}})d^2\Omega=L^2(d\chi^2+sin^2\chi d^2\Omega),\eqno{(89)}$$ with $\chi={{R}\over{L}}\in[0,\pi),$ is the  square lenght element of a 3-sphere $S^3_L$ of radius $L$ whose embedding in $\R^4$ is $L(sin\chi sin\theta cos\varphi,sin\chi sin\theta sin\varphi,sin\chi cos\theta, cos\chi)\in S^3_L\subset\R^4$ with $\chi=0$ corresponding to the north pole $N=(0,0,0,L)$ and $\chi=\pi$ corresponding to the south pole $S=(0,0,0,-L)$. 

\

Note: The metric (88)-(89) can be given an {\it abstract dimensionless} form defining $$\sigma={{s}\over{L}}, \ \ \hat{t}={{cT}\over{L}}\in (-\pi,\pi), \ \ \hat{l}={{l}\over{L}}\in [0,\pi), \eqno{(90)}$$ to obtain $$d\sigma^2={{1}\over{(cos \ \hat{t})+cos \ \chi)^2}}(d\hat{t}^2-d\hat{l}^2), \ \ d\hat{l}^2=d\chi^2+sin^2\chi d^2\Omega.\eqno{(91)}$$ We have an open and bounded space with light cones at $45^\circ$ and $135^\circ$. See Figure 8. 

\

\centerline{\epsfxsize=36ex\epsfbox{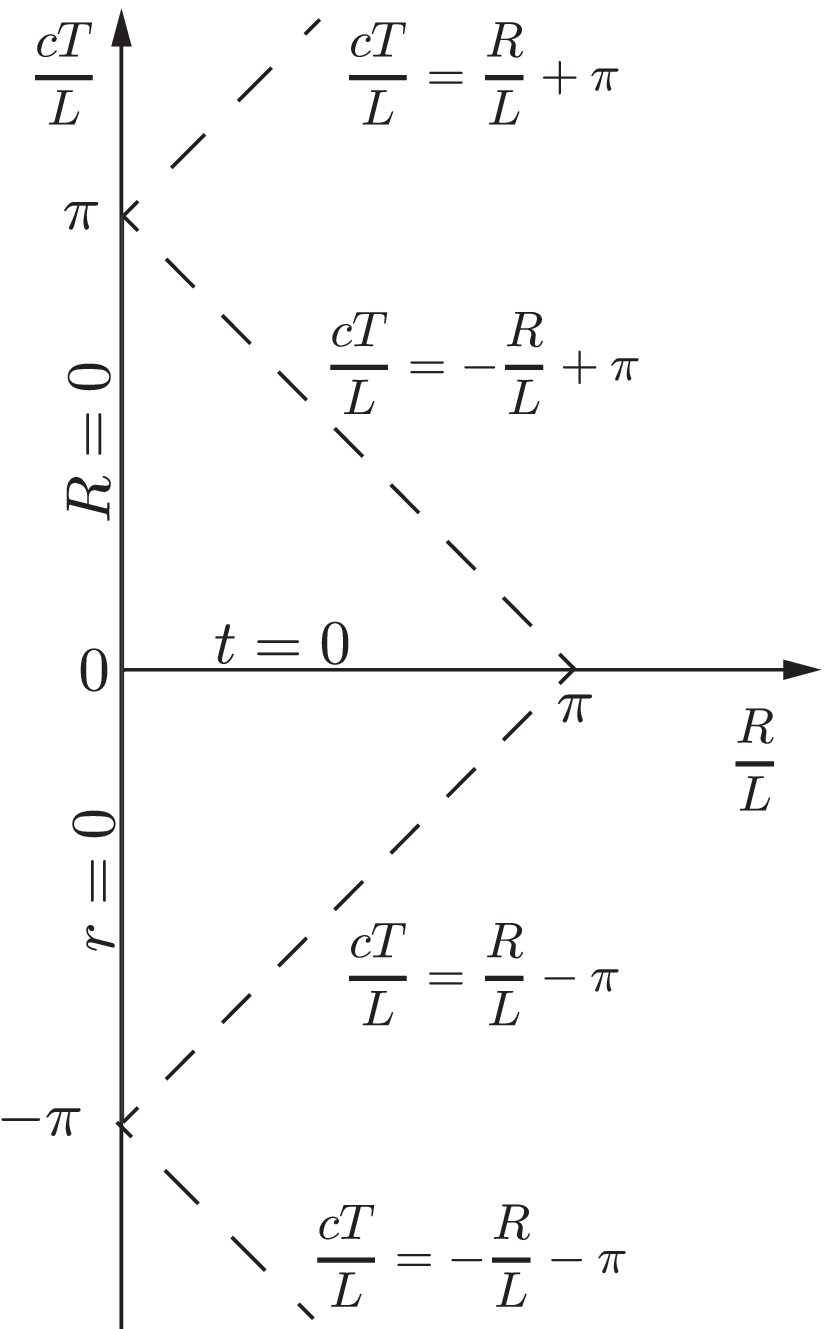}} 

\

\centerline{Figure 8. Open bounded Minkowski space with cartesian coordinates}

\

(v) Conformal transformation: Multiplying $ds^2$ (or $d\sigma^2$) by $(cos({{cT}\over{L}})+cos({{R}\over{L}}))^2$ (or $(cos \ \hat{t})+cos \ \chi)^2$) one obtains the metric 

\

$$d\tilde{s}^2=c^2dT^2-(dR^2+L^2sin^2\chi d^2\Omega)\eqno{(92)}$$ or $$d\tilde{\sigma}^2=d\hat{t}^2-d\hat{l}^2,\eqno{(93)}$$ with the same light cone structure. We obtain the conformal space of Minkowski 4-dimensional space with the topology of an open interval times a 3-sphere i.e. $$conf(Mink^4)\cong(-\pi L,\pi L)\times S^3_L \eqno{(94)}$$ or $$conf(Mink^4)\cong (-\pi,\pi)\times S^3, \eqno{(95)}$$ where $S^3$ is the 3-sphere with unit radius.

\

(vi) Compactification: We add the boundaries (conformal infinity): 

\

$J^+\cong\R\times S^2_L$: future null infinity,

\

$J^-\cong\R\times S^2_L$: past null infinity,

\

$\iota^+\cong\{pt.\}$: future timelike infinity, $(cT\vert_{\iota^+},R\vert_{\iota^+})=(\pi L,0)$,

\

$\iota^-\cong\{pt.\}$: past timelike infinity, $(cT\vert_{\iota^-},R\vert_{\iota^-})=(-\pi L,0)$,

\

and

\

$\iota^0\cong\{pt.\}$: spatial infinity, $(cT\vert_{\iota^0},R\vert_{\iota^0})=(0,\pi L)$, $$\eqno{(96)}$$ obtaining the Penrose space of the 4-dimensional Minkowski space with the topology of a closed interval times a 3-sphere: $$Penr^4\equiv Penr(Mink^4)\cong[-\pi L,\pi L]\times S^3_L,\eqno{(97)}$$ or, without dimensions, $$Penr^4\cong [-\pi,\pi]\times S^3,$$ which is a compact manifold with boundary, see Figure 9. 

\

\centerline{\epsfxsize=40ex\epsfbox{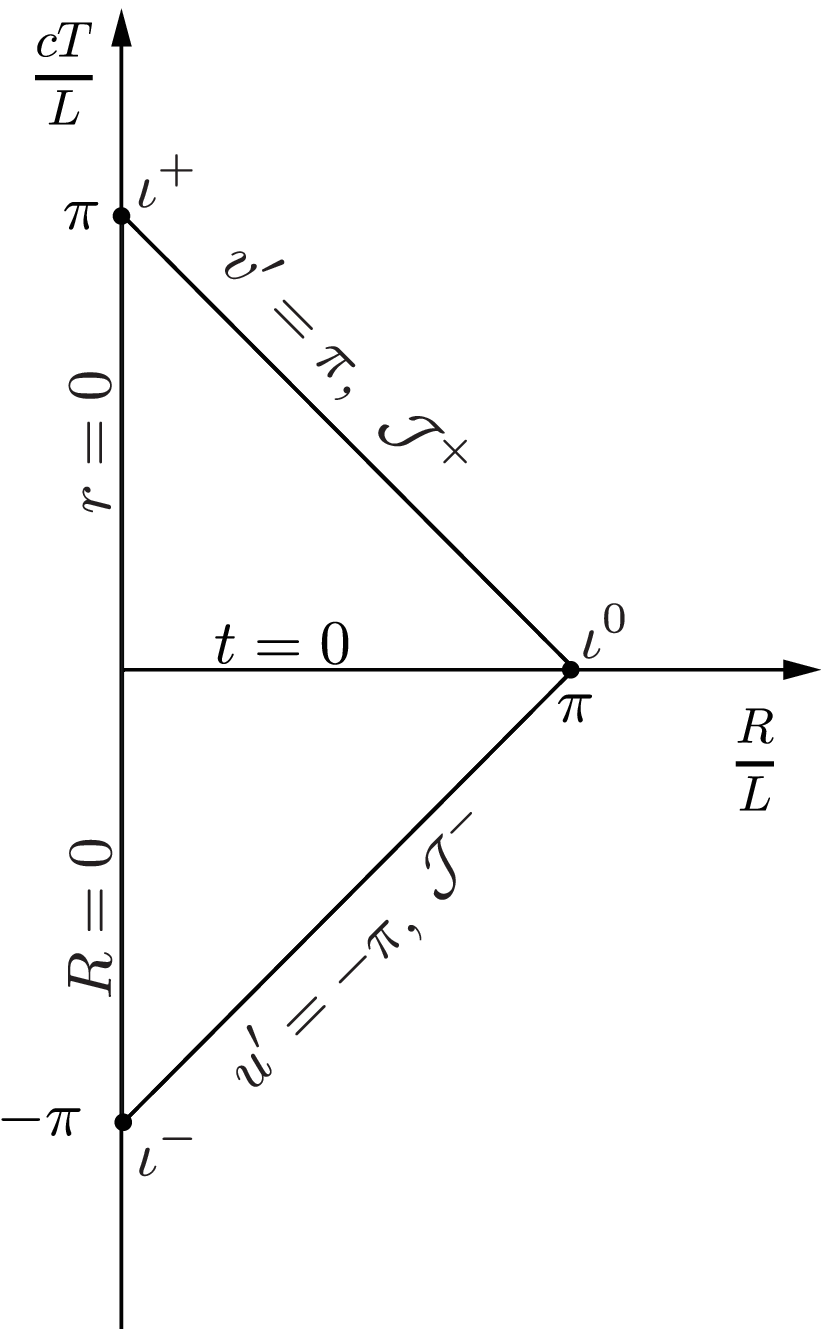}} 

\

\centerline{Figure 9. Penrose space of Minkowski space}

\

Each point in the ``triangle" is a 2-sphere with radius $sin\chi$; the 2-spheres at $\iota^{\pm}$, $\iota^0$, and $\chi={{R}\over{L}}=0$ degenerate into points since $sin\chi=0$ for $R=0$ or $R=\pi L$. Since $vol(S^3_L)=2\pi^2L^3$ we have $$vol(Penr^4)=4\pi^3L^4,\eqno{(98)}$$ with constant scalar curvature $${\cal R}(Penr^4)={\cal R}(S^3_L)={{6}\over{L^2}},\eqno{(99)}$$ respectively given by $$4\pi^3 \ \ and \ \ 6 \eqno{(100)}$$ in the dimensionless coordinates $\hat{t}$ and $\hat{l}$. Though no physical meaning is attributed to these volume and curvature, they are computed to characterize more completely the $Penr^4$ manifold.

\

Note: From (80), (83), and (86),

$$(cT,R)=L(tg^{-1}({{ct+r}\over{L}})+tg^{-1}({{ct-r}\over{L}}),tg^{-1}({{ct+r}\over{L}})-tg^{-1}({{ct-r}\over{L}})).\eqno{(101)}$$

\

{\it Geodesics, timelike, and spacelike curves}

\

It is easy to verify that all radial timelike geodesics begin at $\iota^-$ and end at $\iota^+$, and that all radial spacelike geodesics end at $\iota^0$. 

\

Let $$\gamma: \ \ ct=\alpha r+\beta, \eqno{(102)}$$ with $\alpha$ and $\beta$ constants, be a radial timelike geodesic (straight line); then $v={{dr}\over{dt}}={{c}\over{\alpha}}<c$ and so $\alpha >1$; $t(r=0)={{\beta}\over{c}}$, and $r=\alpha^{-1}(ct-\beta)$. As $t\to\ +\infty$, $r\to +\infty$ with $$u=ct-r=ct-\alpha^{-1}(ct-\beta)=(1-\alpha^{-1})ct+\alpha^{-1}\beta\to\ +\infty\Longrightarrow u\prime\to\pi$$ and $$v=ct+r=ct+\alpha^{-1}(ct-\beta)=(1+\alpha^{-1})ct-\alpha^{-1}\beta\to\ +\infty\Longrightarrow v\prime\to\pi$$ and so $$(cT,R)\to (\pi L,0)=\iota^+, \ t\to+\infty.\eqno{(103)}$$ For $ct=\beta$, $r=0$; then $u=v=\beta$ and so $u^\prime=v^\prime=2tg^{-1}({{\beta}\over{L}})$ and therefore $(cT,R)=(2Ltg^{-1}({{\beta}\over{L}}),0)$. 

\

To consider ``the other side" of $r=0$, we formally extend the domain of $r$ to $r<0$ and allow for $R$ to be in the interval $(-\pi L,\pi L)$. As $t\to -\infty$, $$u=(1-\alpha^{-1})t+\alpha^{-1}\beta\to -\infty\Longrightarrow u^\prime\to -\pi$$ and $$v=(1+\alpha^{-1})t+\alpha^{-1}\beta\to -\infty\Longrightarrow v^\prime\to -\pi$$ and so $$(cT,R)\to (-\pi L,0)=\iota^-, \ t\to-\infty.\eqno{(104)}$$ Inserting (102) into (101) one obtains $$(cT,R)=L(tg^{-1}({{(\alpha+1)r+\beta}\over{L}})+tg^{-1}({{(\alpha-1)r+\beta}\over{L}}),tg^{-1}({{(\alpha+1)r+\beta}\over{L}})-tg^{-1}({{(\alpha-1)r+\beta}\over{L}})).$$ It is then easy to see that $R<0$ if and only if $r<0$. In fact, $R<0\Longleftrightarrow tg^{-1}({{(\alpha+1)r-\vert\beta\vert}\over{L}})<tg^{-1}({{(\alpha-1)r-\vert\beta\vert}\over{L}})$

\

$\Longleftrightarrow (\alpha+1)r-\vert\beta\vert<(\alpha-1)r-\vert\beta\vert\Longleftrightarrow 2r<0$. 

\

For $ct=0$, $r=-{{\beta}\over{\alpha}}={{\vert\beta\vert}\over{\alpha}}$ for $\beta<0$; then $u=-v={{\beta}\over{\alpha}}\Longrightarrow u^\prime=-v^\prime=-2tg^{-1}({{\vert\beta\vert}\over{\alpha}})\Longrightarrow (cT,R)=(0,2Ltg^{-1}({{\vert\beta\vert}\over{\alpha}}))$. 

\

This is illustrated in Figure 10, where $\gamma$, in Penrose space, is $\gamma^\prime$. 

\

\centerline{\epsfxsize=100ex\epsfbox{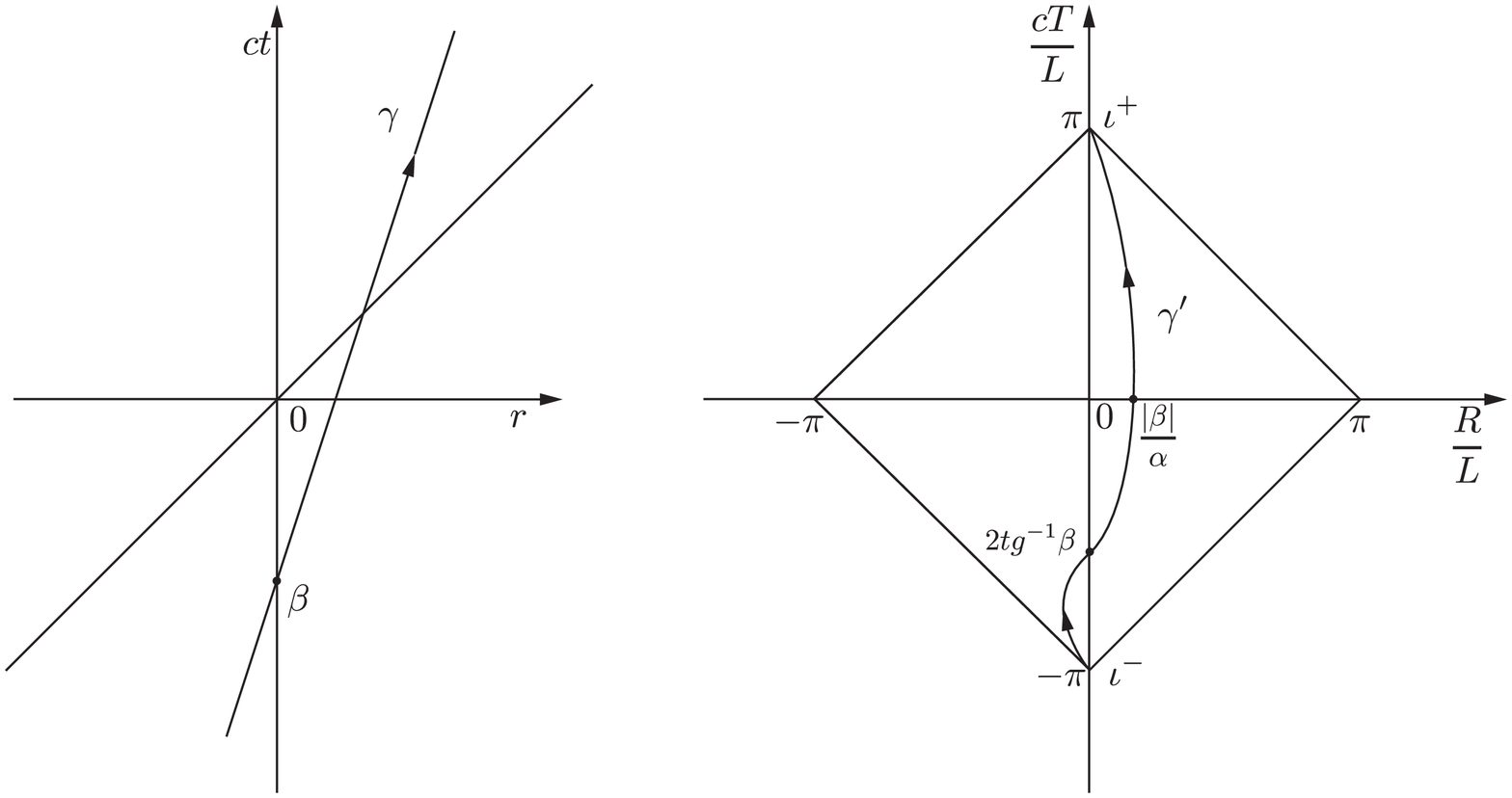}} 

\

\centerline{Figure 10. Timelike radial geodesics in $Mink^4$}

\

Let $$\delta: \ \ ct=\alpha r+\beta, \eqno{(105)}$$ be a radial spacelike geodesic (non physical) and so $\alpha<1$. As before, as $t\to +\infty$: $$u=(1-\alpha^{-1})ct+\alpha^{-1}\beta\to-\infty\Longrightarrow u^\prime\to -\pi$$ and $$v=(1+\alpha^{-1})ct-\alpha^{-1}\beta\to+\infty\Longrightarrow v^\prime\to \pi$$ and so $$(cT,R)\to (0,\pi L)=\iota^0, \ \ t\to +\infty.\eqno{(106)}$$ For $R=0$, $(cT,R)=(2Ltg^{-1}({{\beta}\over{L}}),0)=(-2Ltg^{-1}({{\vert\beta\vert}\over{L}}),0)$. 

\

This is plotted in Figure 11, where $\delta$, in Penrose space, is $\delta^\prime$. 

\

\centerline{\epsfxsize=70ex\epsfbox{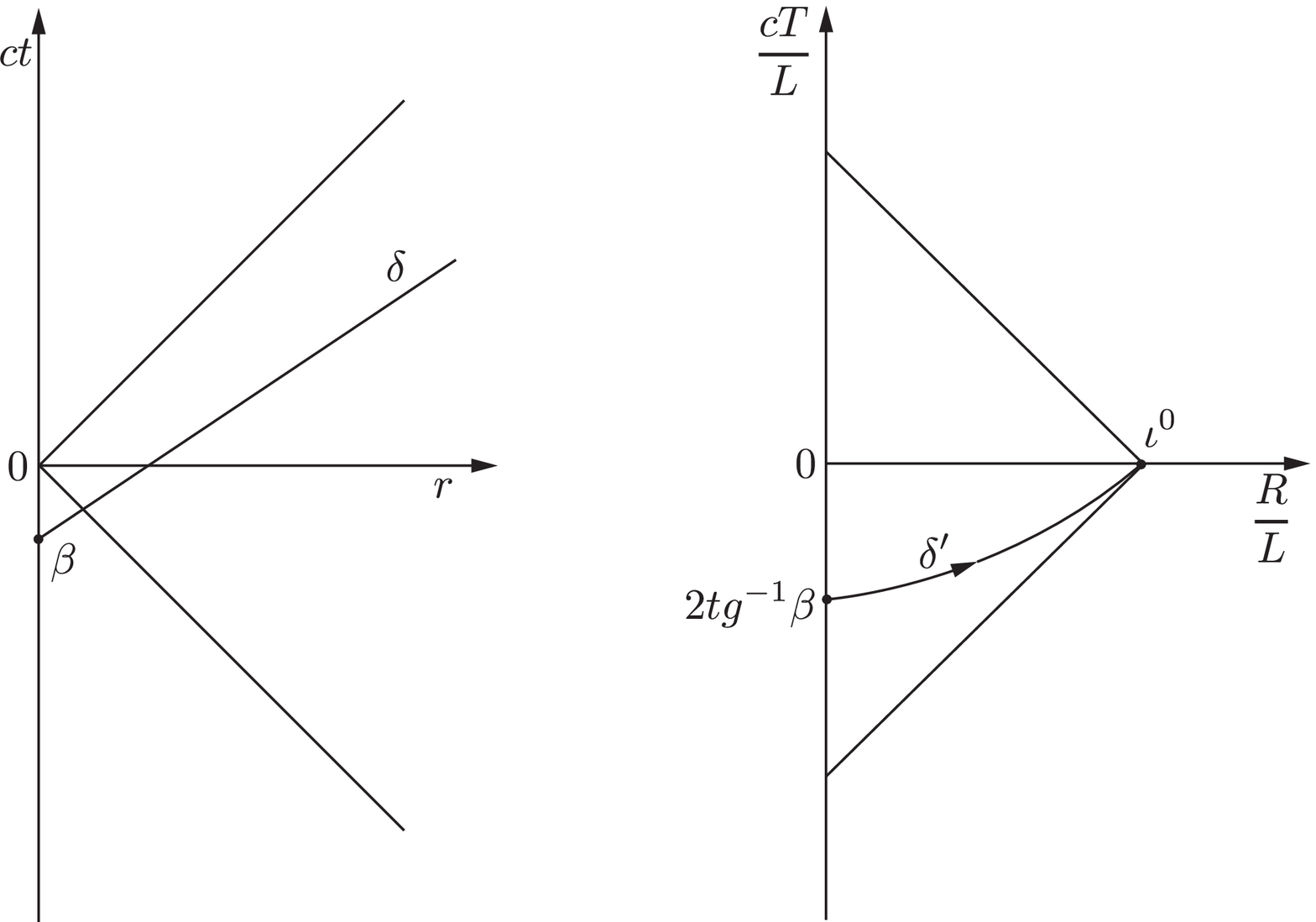}} 

\

\centerline{Figure 11. Spacelike radial geodesics in $Mink^4$}

\

This analysis is easily extended to arbitrary timelike geodesics, not necessarily radial ones: they extend from $\iota^-$ to $\iota^+$ as $t$ goes from $-\infty$ to $+\infty$. Since an arbitrary timelike path can be approximated with arbitrary accuracy with a sequence of null paths, which also are geodesics, the result also holds for arbitrary timelike paths.

\

{\it II.9.2. Rindler space} 

\

In order to describe the Penrose diagram of Rindler space, we restrict ourselves as in section II.1. to 1+1 dimensions, and consider the space $(cT,R)$ with both $cT$and $R$ in the interval $[-\pi,\pi]L$. To the ``square" of Figure 10, we have to add the {\it horizons} $$cT=\pm R\eqno{(107)}$$ or $$\hat{t}=\pm\hat{l}\eqno{(108)}$$ with $\hat{t}$ and $\hat{l}$ in the interval $[-\pi,\pi]$. So, we obtain the space of Figure 12. 

\

\centerline{\epsfxsize=85ex\epsfbox{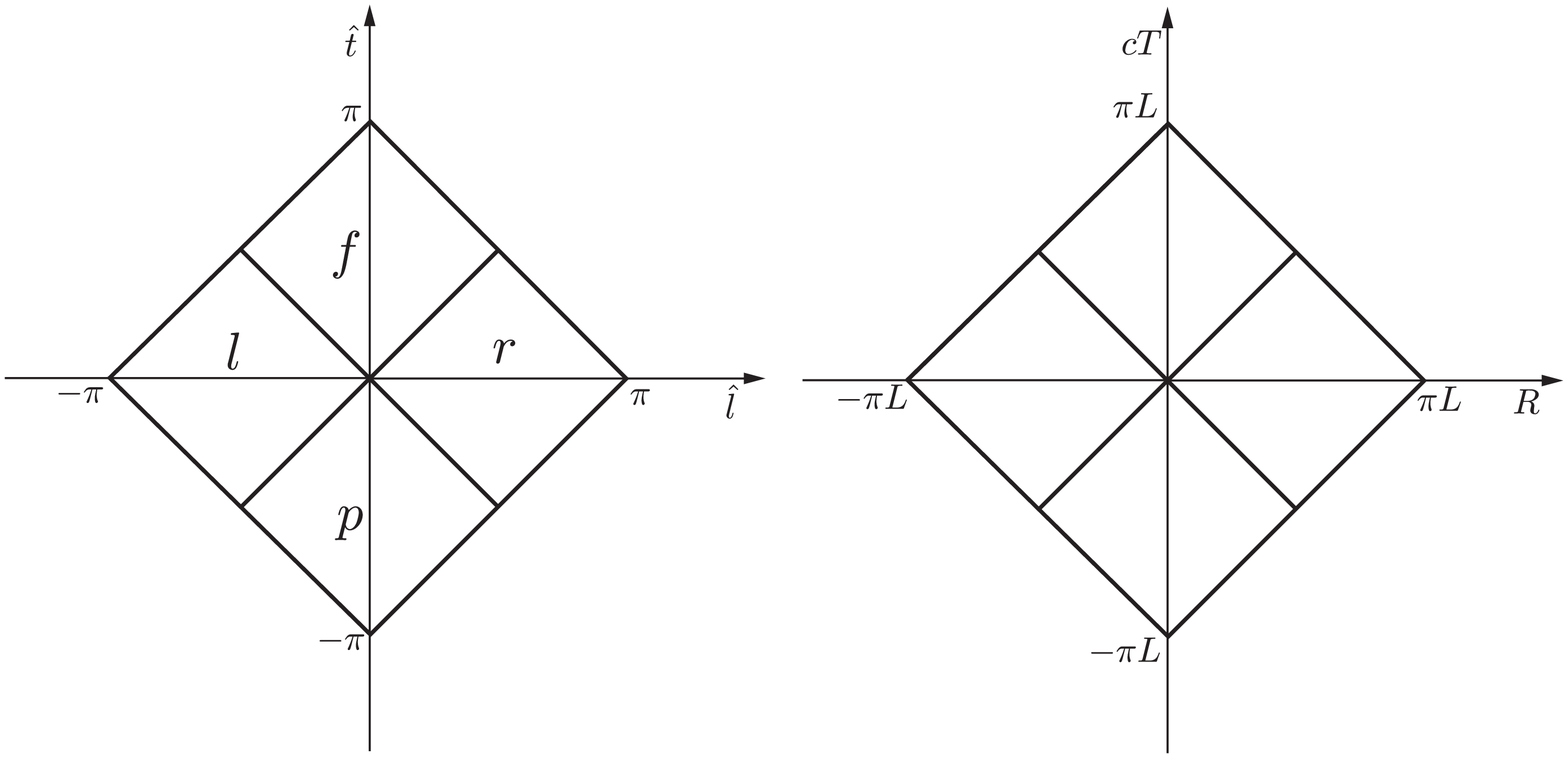}} 

\

\centerline{Figure 12. Penrose space of the four Rindler wedges}

\

The ``diamond" shaped regions r, f, l, and p  respectively correspond to the Rindler wedges $R$, $F$, $L$, and $P$ in Figure 4. 

\

The hyperbolae (accelerated motions in $R$ and $L$) $\alpha$, $\gamma$, $\beta$, and $\delta$ in $R$, $F$, $L$, and $P$ correspond, in the Penrose space, to the curves $\alpha^\prime$, $\gamma^\prime$, $\beta^\prime$, and $\delta^\prime$ in r, f, l, and p respectively. (See Figure 13.) 

\

\centerline{\epsfxsize=120ex\epsfbox{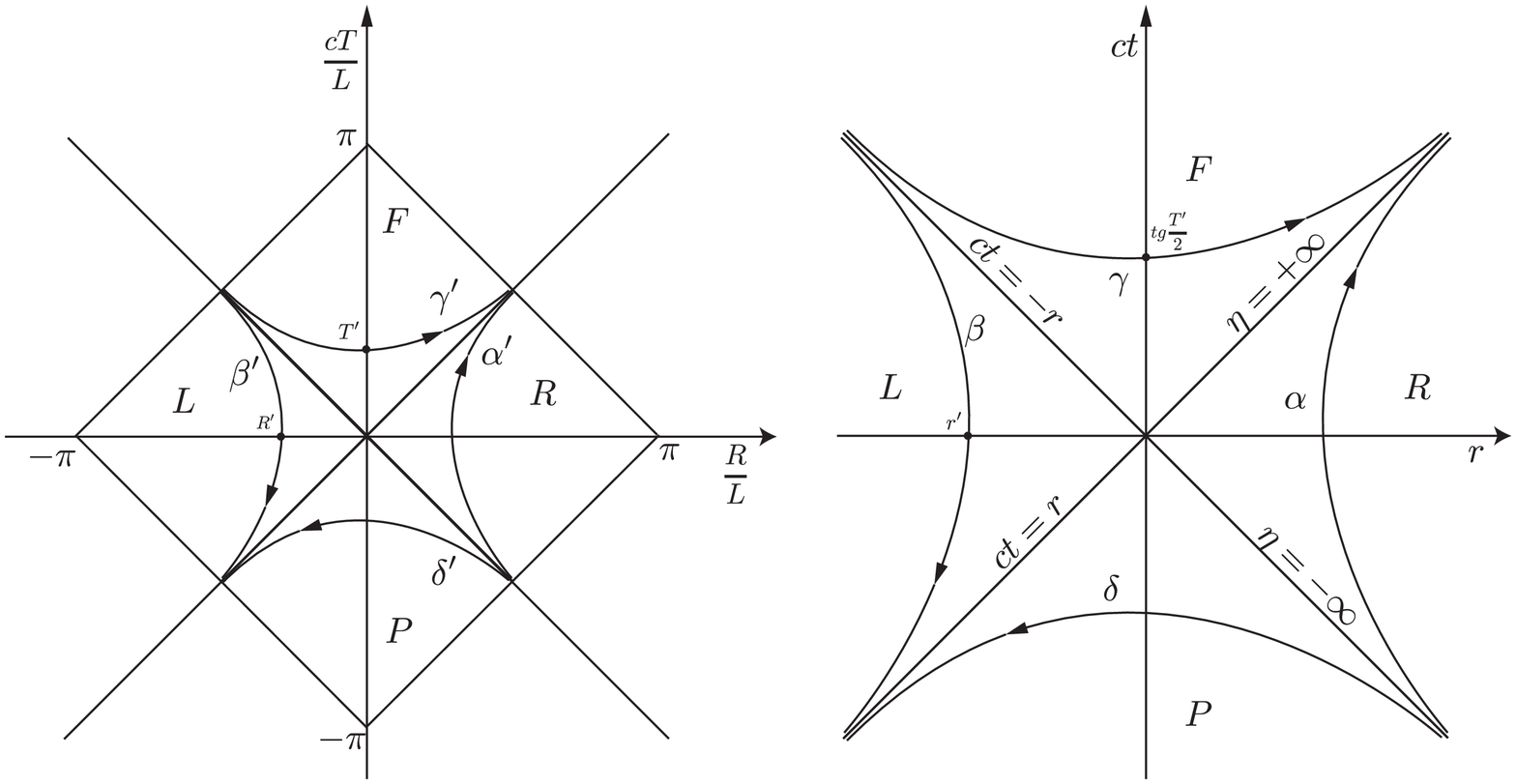}} 

\

\centerline{Figure 13. Hyperbolic trajectories (motions in $R$ and $L$) in the Rindler wedges}

\

Just as, e.g. the hyperbolae $\alpha$'s fill the $R$ wedge, their ``images" $\alpha^\prime$'s fill the diamond r. The analogous happens in the regions f, l, and p.  

\

In r/$R$:

\

Take $(cT,R)=(0,R^\prime)$, with $R^\prime\in(0,\pi L)$; in $(ct,x)$ it corresponds to $(0,Ltg({{R^\prime}\over{2L}}))$ (we used (101) with $t=0$ and $r$ replaced by $x$). So, $0<{{R^\prime}\over{2}}<{{\pi}\over{2}}\Longrightarrow 0<x=tg({{R^\prime}\over{2L}})<+\infty$.

\

$(cT,R)=({{\pi}\over{2}}L,{{\pi}\over{2}}L)=L(tg^{-1}(ct+x)+tg^{-1}(ct-x),tg^{-1}(ct+x)-tg^{-1}(ct-x))\Longrightarrow tg^{-1}(ct-x)=0\Longrightarrow ct=x$: asymptotic to future horizon.

\

$(cT,R)=(-{{\pi}\over{2}}L,{{\pi}\over{2}}L)=L(tg^{-1}(ct+x)+tg^{-1}(ct-x),tg^{-1}(ct+x)-tg^{-1}(ct-x))\Longrightarrow tg^{-1}(ct+x)=0\Longrightarrow ct=-x$: asymptotic to past horizon.

\

For example, to $R={{\pi}\over{2}}$ corresponds $x=Ltg({{\pi}\over{4}})=L$ (=1 in dimensionless coordinates). Obviously, for $x^\prime<x$, $R^\prime<R$. 

\

Then, $$\alpha\leftrightarrow\alpha^\prime.\eqno{(109)}$$ 

\

In l/$L$:

\

Take $(cT,R)=(0,R^\prime)$, with $R^\prime\in(-\pi L,0)$; in $(ct,x)$ it corresponds to $(0,Ltg({{R^\prime}\over{2L}}))$ with $tg({{R^\prime}\over{2L}})\in(-\infty,0)$.

\

$(cT,R)=({{\pi}\over{2}}L,-{{\pi}\over{2}}L)=L(tg^{-1}(ct+x)+tg^{-1}(ct-x),tg^{-1}(ct+x)-tg^{-1}(ct-x))\Longrightarrow tg^{-1}(ct+x)=0\Longrightarrow ct=-x$: asymptotic to past horizon.

\

$(cT,R)=(-{{\pi}\over{2}}L,-{{\pi}\over{2}}L)=L(tg^{-1}(ct+x)+tg^{-1}(ct-x),tg^{-1}(ct+x)-tg^{-1}(ct-x))\Longrightarrow tg^{-1}(ct-x)=0\Longrightarrow ct=x$: asymptotic to future horizon.

\

Then, $$\beta\leftrightarrow\beta^\prime.\eqno{(110)}$$ 

\

In f/$F$:

\

$(cT,R)=({{\pi}\over{2}}L,-{{\pi}\over{2}}L)\Longrightarrow ct=x$: asymptotic to past horizon.

\

$(cT,R)=({{\pi}\over{2}}L,{{\pi}\over{2}}L)\Longrightarrow ct=x$: asymptotic to future horizon.

\

$cT=cT^\prime\in(0,\pi L)$, $R^\prime=0 \Longrightarrow x^\prime=0$ and $0<ct^\prime=Ltg({{cT^\prime}\over{2L}})<+\infty$.

\

Then, $$\gamma\leftrightarrow\gamma^\prime.\eqno{(111)}$$ 

\

In p/$P$:

\

$(cT,R)=(-{{\pi}\over{2}}L,{{\pi}\over{2}}L)\Longrightarrow ct=x$: asymptotic to future horizon.

\

$(cT,R)=(-{{\pi}\over{2}}L,{{\pi}\over{2}}L)\Longrightarrow ct=-x$: asymptotic to past horizon.

\

$cT=cT^{\prime\prime}\in(0,-\pi L)$, $R^{\prime\prime}=0\Longrightarrow x^{\prime\prime}=0$, $0>ct^{\prime\prime}=Ltg({{cT^{\prime\prime}}\over{2L}})>-\infty$.

\

Then, $$\delta\leftrightarrow\delta^\prime.\eqno{(112)}$$ 

\

{\bf III. The Unruh effect.}

\

{\it III.1. Bogoliubov transformations}

\

Let $\phi$ be a free 2-dimensional real massless scalar field and let ${\cal U}=\{u_i\}_{i\in{\cal I}}$ and ${\cal V}=\{v_K\}_{k\in{\cal J}}$ be two complete sets of positive frequency solutions of the Klein-Gordon equation $$g^{\mu\nu}\partial_\mu\partial_\nu\phi=0.\eqno{(113)}$$ (${\cal I}$ and ${\cal J}$ are sets of indices.) The $u_i$'s and $v_K$'s obey the orthonormalization conditions (for simplicity we adopt discrete sets of indices) $$<u_i,u_j>=-<u_i^*,u_j^*>=\delta_{ij}, \ \ <u_i,u_j^*>=0,$$ $$<v_K,v_L>=-<v_K^*,v_L^*>=\delta_{KL}, \ \ <v_K,v_L^*>=0\eqno{(114)}$$ where the complex conjugates $u_j^*$'s and $v_L^*$'s are negative frequency solutions. < , > is the scalar product in the space of the field.

\

Let $a_i$ and $a_i^\dagger$ be the annihilation and creation operators associated to ${\cal U}$ and $b_K$ and $b_K^\dagger$ those associated with ${\cal V}$. They obey the commutation relations $$[a_i,a_j^\dagger]=\delta_{ij}, \ \ [a_i,a_j]=[a_i^\dagger,a_j^\dagger]=0,$$ $$[b_K,b_L^\dagger]=\delta_{KL}, \ \ [b_K,b_L]=[b_K^\dagger,b_L^\dagger]=0.\eqno{(115)}$$

\

The field operator $\phi$ has the expansions $$\phi=\sum_{i\in{\cal I}}(u_ia_i+u_i^*a_i^\dagger)=\sum_{K\in{\cal J}}(v_Kb_K+v_K^*b_K^\dagger).\eqno{(116)}$$ There are normalized vacuum states in the corresponding Fock spaces, $\vert 0>_a$ and $\vert 0>_b$, with ${_a}<0\vert 0>_a={_b}<0\vert 0>_b=1$, satisfying $$a_i\vert 0>_a=0, \ \ b_K\vert 0>_b=0,\eqno{(117)}$$ 1-particle states $$\vert 1_i>_a=a_i^\dagger\vert 0>_a, \ \ \vert 1_K>_b=b_K^\dagger\vert 0>_b, \eqno{(118)}$$ etc.

\

The completeness of ${\cal U}$ allows to express the eigenfunctions $v_K$'s in terms of the $u_i$'s: $$v_K=\sum_{i\in{\cal I}}(\alpha_{Ki}u_i+\beta_{Ki}u_i^*),$$ $$v_K^*=\sum_{i\in{\cal I}}(\alpha_{Ki}^*u_i^*+\beta_{Ki}^*u_i).\eqno{(119)}$$ The quantities $\alpha_{Ki}$ and $\beta_{Ki}$ are the {\it Bogoliubov transformation coefficients} between the complete sets ${\cal U}$ and ${\cal V}$. Notice that a non vanishing $\beta_{Ki}$ gives a negative frequency contribution (that of $u_i^*$) to the positive frequency solution $v_K$. Replacing (119) in (116) and taking into account the completeness of the $u_i$'s, one obtains the transformation of the creation and annihilation operators: $$a_i=\sum_{K\in{\cal J}}(\alpha_{Ki}b_K+\beta_{Ki}^*b_K^\dagger),$$ $$a_i^\dagger=\sum_{K\in{\cal J}}(\alpha_{Ki}^*b_{Ki}^\dagger+\beta_{Ki}b_K).\eqno{(120)}$$ The orthonormality conditions (114) lead to the following relations between the Bogoliubov coefficients: $$\sum_{i\in{\cal I}}(\alpha_{Ki}\alpha_{Li}^*-\beta_{Ki}\beta_{Li}^*)=\delta_{KL},$$ $$\sum_{i\in{\cal I}}(\alpha_{Ki}\beta_{Li}-\beta_{Ki}\alpha_{Li}=0.\eqno{(121)}$$ From the scalar products $<u_i,v_K>=<u_i,\sum_{j\in{\cal I}}\alpha_{Kj}u_j>=\alpha_{Ki}$, $<u_i,v_K^*>=<u_i,\sum_{j\in{\cal I}}\beta_{Kj}^*u_j>=\beta_{Ki}^*$, and the expansion $u_i=\sum_{K\in{\cal J}}(r_{iK}v_K+s_{iK}v_K^*)$, the scalar products $<u_i,v_L>=r_{iL}^*$ and $<u_i,v_L^*>=-s_{iL}^*$ lead to the identifications $r_{iL}^*=\alpha_{Li}$ and $s_{Li}^*=-\beta_{Li}^*$ and therefore to the inverse transformations $$u_i=\sum_{K\in{\cal J}}(\alpha_{Ki}^*v_K-\beta_{Ki}v_K^*),$$ $$u_i^*=\sum_{K\in{\cal J}}(\alpha_{Ki}v_K^*-\beta_{Ki}^*v_K)\eqno{(122)}$$ and $$b_K=\sum_{i\in{\cal I}}(\alpha_{Ki}a_i-\beta_{Ki}^*a_i^\dagger),$$ $$b_K^\dagger=\sum_{i\in{\cal I}}(\alpha_{Ki}^*a_i^\dagger-\beta_{Ki}a_i),\eqno{(123)}$$ with the relations $$\sum_{K\in{\cal J}}(\alpha_{Ki}\alpha_{Kj}^*-\beta_{Ki}^*\beta_{Kj})=\delta_{ij},$$ $$\sum_{K\in{\cal J}}(\alpha_{Ki}\beta_{Kj}^*+\beta_{Ki}^*\alpha_{Kj})=0.\eqno{(124)}$$

\

At this point, we ask the following question: Which is the average value of the occupation number operator of the $b$ particles, $$n_K=b_K^\dagger b_K,\eqno{(125)}$$ in the vacuum state of the $a$ particles, $\vert 0>_a$? Since $\vert 0>_a$ is defined by the condition (117), we have $${_a}<0\vert n_K\vert 0>_a=\sum_{i,j\in{\cal I}}{_a}<0\vert(\alpha_{Ki}^*a_i^\dagger-\beta_{Ki}a_i)(\alpha_{Kj}a_j-\beta_{Kj}^*a_j^\dagger)\vert 0>_a=\sum_{i,j\in{\cal I}}\beta_{Ki}\beta_{Kj}^* \ {_a}<0\vert a_ia_j^\dagger\vert 0>_a$$ $$=\sum_{i,j\in{\cal I}}\beta_{Ki}\beta_{Kj}^*\delta_{ij}=\sum_{i\in{\cal I}}\vert\beta_{Ki}\vert^2\eqno{(126)}$$ with $\sum_{i\in{\cal I}}\vert\beta_{Ki}\vert^2\neq 0$ if at least one of the $\beta_{Ki}$'s is different from zero. So, the observer using the $b$ basis can see particles in the $a$ vacuum. The necessary and sufficient condition for that, is that at least one of the positive frequency modes in ${\cal V}$ had a non vanishing contribution of a negative frequency mode in ${\cal U}$. 

\

{\it III.2. Unruh effect}

\

To the positive (negative) frequency Minkowski modes $\varphi_k^{(a)}$ (${\varphi_k^{(a)}}^*$), $a=1,2$, of equations (76)-(77), correspond the Minkowski operators ${a_k^{(a)M}}$ (${a_k^{(a)M}}^\dagger$), $a=1,2$, which obey, in particular, $${a_k^{(a)M}}\vert 0>_M=0.\eqno{(127)}$$ To the positive (negative) frequency Rindler modes $\phi_k^R$ (${\phi_k^R}^\dagger$) and  $\phi_k^L$ (${\phi_k^L}^\dagger$) in the right and left Rindler wedges, correspond, respectively, the operators $b_k^R$ (${b_k^R}^\dagger$) and $b_k^L$ (${b_k^L}^\dagger$) which obey, in particular, $$b_k^R\vert 0>_R=b_k^L\vert 0>_L=0. \eqno{(128)}$$ If, in the general theory of the Bogoliubov coefficients we identify the $b_k^R$'s and $b_k^L$'s with the $b_K$'s, and the $a_k^{(a)M}$'s, $a=1,2$, with the $a_i$'s or, equivalently, the $\phi_k^R$'s and $\phi_k^L$'s with the $v_K$'s and the $\varphi_k^{(a)}$'s, $a=1,2$, with the $u_i$'s, from the expansions (74) and (75), and (122), we can identify the $\alpha_{Ki}$'s and the $\beta_{Ki}$'s: $$\alpha_{Ki}^*={{e^{\pi\omega c/2a}}\over{\sqrt{2Sh({{\pi\omega c}\over{a}})}}}, \ \ \beta_{Ki}=-{{e^{-\pi\omega c/2a}}\over{\sqrt{2Sh({{\pi\omega c}\over{a}})}}}.\eqno{(129)}$$ Then, $$b_k^R={{e^{\pi\omega c/2a}}\over{\sqrt{2Sh({{\pi\omega c}\over{a}})}}}a_k^{(1)M}-{{e^{-\pi\omega c/2a}}\over{\sqrt{2Sh({{\pi\omega c}\over{a}})}}}{a_{-k}^{(2)M}}^\dagger,$$ $${b_k^R}^\dagger={{e^{\pi\omega c/2a}}\over{\sqrt{2Sh({{\pi\omega c}\over{a}})}}}{a_k^{(1)M}}^\dagger-{{e^{-\pi\omega c/2a}}\over{\sqrt{2Sh({{\pi\omega c}\over{a}})}}}a_{-k}^{(2)M}.\eqno{(130)}$$ So, with $$n_k^R={b_k^R}^\dagger b_k^R, \eqno{(131)}$$ $${_ M}<0\vert n_k^R\vert 0>_M={{(e^{\pi\omega c/2a})^2}\over{\sqrt{2Sh({{\pi\omega c}\over{a}})}}}\times\delta(0)={{1}\over{{e^{{2\pi\omega c}\over{a}}}-1}}\times\delta(0)={{1}\over{e^{{\omega c}\over{{a/2\pi}}}-1}}\times\delta(0).\eqno{(132)}$$ $\omega$ is the frequency of the Rindler mode in the coordinate system $(\lambda,\xi)$ (equation (64)); equation (19) gives the relation between $\lambda$ and proper acceleration $\alpha$ and proper time $\tau$ of a Rindler observer. For a period $T$ (with $\omega T=2\pi$) one has $aT=\alpha T_{pr}$ i.e. ${{a}\over{\omega}}={{\alpha}\over{\omega_{pr}}}$, then $${{\omega}\over{a}}={{\omega_{pr}}\over{\alpha}}\eqno{(133)}$$ and therefore $${_M}<0\vert n_k^R\vert 0>_M={{1}\over{e^{{\omega_{pr} c}\over{{a/2\pi}}}-1}}\times\delta(0).\eqno{(134)}$$ The infinite factor $\delta(0)$ comes from the continuous commutator $[a_l^{(2)M},{a_m^{(2)M}}^\dagger]=\delta(l-m)$. 

\

The r.h.s. of (134) is a thermal distribution (a Bose-Einstein distribution) with {\it absolute temperature} $${\cal T}={{\hbar\alpha}\over{2\pi k_Bc}}\eqno{(135)}$$ where $k_B$ is the Boltzmann constant. In fact, ${{\omega_{pr}c}\over{\alpha/2\pi}}={{\hbar\omega_{pr}c}\over{\hbar\alpha/2\pi}}={{\hbar\omega_{pr}}\over{\hbar\alpha/2\pi c}}\equiv\epsilon_{pr}/k_B{\cal T}$ i.e. $${_M}<0\vert n_k^R\vert 0>_M={{1}\over{e^{{\epsilon_{pr}}\over{k_B{\cal T}}}-1}}\times\delta(0).\eqno{(136)}$$ This is the Unruh effect.

\

Of course, the same result is obtained in the left Rindler wedge: with the same identifications as before, $$b_k^L={{e^{\pi\omega c/2a}}\over{\sqrt{2Sh({{\pi\omega c}\over{a}})}}}a_k^{(2)M}+{{e^{-\pi\omega c/2a}}\over{\sqrt{2Sh({{\pi\omega c}\over{a}})}}}{a_{-k}^{(1)M}}^\dagger,$$ $${b_k^L}^\dagger={{e^{\pi\omega c/2a}}\over{\sqrt{2Sh({{\pi\omega c}\over{a}})}}}{a_k^{(2)M}}^\dagger+{{e^{-\pi\omega c/2a}}\over{\sqrt{2Sh({{\pi\omega c}\over{a}})}}}a_{-k}^{(1)M},\eqno{(137)}$$ we obtain $${_M}<0\vert n_k^L\vert 0>_M={_M}<0\vert n_k^R\vert 0>_M.\eqno{(138)}$$

\

{\it III.3. Unruh effect: a more direct calculation (Lee, 1986)}

\

Consider again a free real massless scalar field $\phi$ in Minkowski space $$\phi(t,x)=\int_{-\infty}^{+\infty}{{dk}\over{2\pi\sqrt{2\vert k\vert}}}(a_ke^{-i(\omega_kt-kx)}+a_k^\dagger e^{i(\omega_kt-kx)})\eqno{(139)}$$ and in the Rindler wedge $R$ $$\phi(\lambda,\xi)=\int_{-\infty}^{+\infty}{{dk}\over{2\pi\sqrt{2\vert l\vert}}}(\alpha_le^{-i(\omega_l\lambda-l\xi)}+\alpha_l^\dagger e^{i(\omega_l\lambda-l\xi)}),\eqno{(140)}$$ where 

\

$\omega_k=c\vert k\vert$, $\omega_l=c\vert l\vert$, $[k]=[l]=[L]^{-1}$, $[a_k,a_{k^\prime}^{\dagger}]=2\pi\delta(k-k^\prime)$, $[a_k,a_{k^\prime}]=[a_k^\dagger,a_{k^\prime}^{\dagger}]=0,$ $[\alpha_l,\alpha_{l^\prime}^{\dagger}]=2\pi\delta(l-l^\prime)$, $$[\alpha_l,\alpha_{l^\prime}]=[\alpha_l^\dagger,\alpha_{l^\prime}^{\dagger}]=0.\eqno{(141)}$$ 

\

At $t=0$, $$\phi(0,x)=\int_{-\infty}^{+\infty}{{dk}\over{2\pi\sqrt{2\vert k\vert}}}(a_ke^{ikx}+a_k^\dagger e^{-ikx})=\int_{-\infty}^{+\infty}{{dk}\over{2\pi\sqrt{2\vert k\vert}}}(a_ke^{i{{k}\over{a}}c^2e^{a\xi/c^2}}+a_k^\dagger e^{-i{{k}\over{a}}c^2e^{a\xi/c^2}})=\phi(0,\xi),\eqno{(142)}$$ where we used (16) and the fact that $\lambda=0\Longleftrightarrow t=0$. On the other hand, the Fourier transform of (140) is $$\int_{-\infty}^{+\infty}d\xi e^{-il^\prime\xi}\phi(\lambda,\xi)=\int_{-\infty}^{+\infty}{{dl}\over{2\pi\sqrt{2\vert l\vert}}}(\alpha_le^{-i\vert l\vert\lambda}\int_{-\infty}^{+\infty}d\xi e^{i(l-l^\prime)\xi}+\alpha_l^\dagger e^{i\vert l\vert\lambda}\int_{-\infty}^{+\infty}d\xi e^{-i(l-l^\prime)\xi})$$ $$=\int_{-\infty}^{+\infty}{{dl}\over{2\pi\sqrt{2\vert l\vert}}}(\alpha_le^{-i\vert l^\prime\vert\lambda}2\pi\delta(l-l^\prime)+\alpha_l^\dagger e^{i\vert l\vert\lambda}2\pi\delta(l+l^\prime)={{1}\over{\sqrt{2\vert l^\prime\vert}}}(\alpha_{l^\prime}e^{-i\vert l^\prime\vert\lambda}+\alpha_{-l^\prime}^\dagger e^{i\vert l^\prime\vert\lambda}),$$ which, for $\lambda=0$ and using (142) gives $$\alpha_l+\alpha_{-l}^\dagger=\sqrt{2\vert l\vert}\int_{-\infty}^{+\infty}d\xi\phi(0,\xi)e^{-il\xi}$$ $$=\int_{-\infty}^{+\infty}{{dk}\over{2\pi}}\sqrt{{{\vert l\vert}\over{\vert k\vert}}}(a_k\int_{-\infty}^{+\infty}d\xi e^{i(c^2{{k}\over{a}}e^{a\xi/c^2}-l\xi)}+a_k^\dagger\int_{-\infty}^{+\infty}d\xi e^{-i(c^2{{k}\over{a}}e^{a\xi/c^2}+l\xi)}).\eqno{(143)}$$ From (140), $${{\partial}\over{\partial\lambda}}\phi(\lambda,\xi)\vert_{\lambda=0}=-i\int_{-\infty}^{+\infty}{{dl}\over{2\pi}}\sqrt{{{\vert l\vert}\over{2}}}(\alpha_le^{il\xi}-\alpha_l^\dagger e^{-il\xi})$$ and so $$\int_{-\infty}^{+\infty}d\xi e^{-il^\prime\xi}{{\partial}\over{\partial\lambda}}\phi(\lambda,\xi)\vert_{\lambda=0}=-i\sqrt{{{\vert l^\prime\vert}\over{2}}}(\alpha_{l^\prime}-\alpha_{-l^\prime}^\dagger)$$ i.e. $$\alpha_l-\alpha_{-l}^\dagger =\int_{-\infty}^{+\infty}{{dk}\over{2\pi}}\sqrt{{{\vert k\vert}\over{\vert l\vert}}}(a_k\int_{-\infty}^{+\infty}d\xi e^{a\xi/c^2}e^{i(c^2{{k}\over{a}}e^{a\xi/c^2}-l\xi)}-a_k^\dagger\int_{-\infty}^{+\infty}d\xi e^{a\xi/c^2}e^{-i(c^2{{k}\over{a}}e^{a\xi/c^2}+l\xi)}).\eqno{(144)}$$ Defining $$<l,k>:=\int_{-\infty}^{+\infty}d\xi e^{i(c^2{{k}\over{a}}e^{a\xi/c^2}-l\xi)},\eqno{(145)}$$ and $$(l,k)=\int_{-\infty}^{+\infty}d\xi e^{a\xi/c^2}e^{i(c^2{{k}\over{a}}e^{a\xi/c^2}-l\xi)},\eqno{(146)}$$ with $$<l,k>=<-l,-k>^*, \ \ <l,-k>=<-l,k>^*,\eqno{(147)}$$ and $$(l,k)=-i{{a}\over{c^2}}{{\partial}\over{\partial k}}<l,k>, \ \ (l,-k)=i{{a}\over{c^2}}{{\partial}\over{\partial k}}<l,-k>,\eqno{(148)}$$ ($[<l,k>]=[(l,k)]=[L]$) one obtains $$\alpha_l={{1}\over{2}}\int_{-\infty}^{+\infty}{{dk}\over{2\pi}}(a_k(\sqrt{{{\vert l\vert}\over{\vert k\vert}}}-i{{a}\over{c^2}}\sqrt{{{\vert k\vert}\over{\vert l\vert}}}{{\partial}\over{\partial k}})<l,k>+a_k^\dagger (\sqrt{{{\vert l\vert}\over{\vert k\vert}}}-i{{a}\over{c^2}}\sqrt{{{\vert k\vert}\over{\vert l\vert}}}{{\partial}\over{\partial k}})<l,-k>)),$$ $$\alpha_{-l}^\dagger ={{1}\over{2}}\int_{-\infty}^{+\infty}{{dk}\over{2\pi}}(a_k(\sqrt{{{\vert l\vert}\over{\vert k\vert}}}+i{{a}\over{c^2}}\sqrt{{{\vert k\vert}\over{\vert l\vert}}}{{\partial}\over{\partial k}})<l,k>+a_k^\dagger (\sqrt{{{\vert l\vert}\over{\vert k\vert}}}+i{{a}\over{c^2}}\sqrt{{{\vert k\vert}\over{\vert l\vert}}}{{\partial}\over{\partial k}})<l,-k>)).\eqno{(149)}$$ For $<l,k>$ one has: $$<l,k>=I_1+I_2$$ with $$I_1=\int_{0}^{+\infty} d\xi e^{i(c^2{{k}\over{a}}e^{a\xi/c^2}-l\xi)}={{c^2}\over{a}}\int_{0}^{+\infty}dx e^{ic^2({{k}\over{a}}e^x-{{l}\over{a}}x)}={{c^2}\over{a}}(-i{{k}\over{a}}c^2)^{ic^2{{l}\over{a}}}\Gamma(-ic^2{{l}\over{a}},-ic^2{{k}\over{a}}),\eqno{(150)}$$ $$I_2=\int_{0}^{+\infty} d\xi e^{i(c^2{{k}\over{a}}e^{a\xi/c^2}+l\xi)}={{c^2}\over{a}}\int_{0}^{+\infty}dx e^{ic^2({{k}\over{a}}e^{-x}+{{l}\over{a}}x)}={{c^2}\over{a}}(-i{{k}\over{a}}c^2)^{ic^2{{l}\over{a}}}\gamma(-ic^2{{l}\over{a}},-ic^2{{k}\over{a}}),\eqno{(151)}$$ where $\Gamma(\alpha,x)=\int_{x}^{+\infty}dte^{-t}t^{\alpha-1}$ (Gradshtein and Ryzhik, 1980; 8.350.2, p. 940) and $\gamma(\alpha,x)=\int_{0}^{x}dte^{-t}t^{\alpha-1}$ (Gradshtein and Ryzhik, 1980; 8.350.1, p. 940) are the incomplete $\Gamma$-functions. Using $\Gamma(\alpha,x)+\gamma(\alpha,x)=\Gamma(\alpha)$ (Gradshtein and Ryzhik, 1980; 8.356.3, p. 942), for $l,k>0$ one obtains $$<l,k>=k^{ic^2{{l}\over{a}}}f(l),\eqno{(152)}$$ with $$f(l)={{1}\over{c^2}}a^{-ic^2{{l}\over{a}}-1}e^{{{\pi lc^2}\over{2a}}}\Gamma(-ic^2{{l}\over{a}}).\eqno{(153)}$$ Then, from (148), $$(l,k)={{l}\over{k}}<l,k>.\eqno{(154)}$$ Replacing these results in (149), we obtain $$\alpha_l=\int_{0}^{+\infty}{{dk}\over{2\pi}}\sqrt{{{l}\over{k}}}(a_k<l,k>+a_k^\dagger <l,-k>), \ \ l>0,$$ $$\alpha_l=\int_{0}^{+\infty}{{dk}\over{2\pi}}\sqrt{{{-l}\over{k}}}(a_{-k}^\dagger <l,k>+a_{-k}<l,-k>), \ \ l<0,\eqno{(155)}$$ and $$\alpha_l^\dagger =\int_{0}^{+\infty}{{dk}\over{2\pi}}\sqrt{{{l}\over{k}}}(a_k^\dagger <-l,-k>+a_k<-l,k>), \ \ l>0,$$ $$\alpha_l^\dagger =\int_{0}^{+\infty}{{dk}\over{2\pi}}\sqrt{{{-l}\over{k}}}(a_{-k} <-l,-k>+a_{-k}^\dagger <-l,k>), \ \ l<0,\eqno{(156)}$$ where the Bogoliubov coefficients between the creation and annihilation operators in Minkowski and Rindler spaces as in (120) are given by $<l,k>$, $<l,-k>$, etc. 

\

The calculation of $_M<0\vert\alpha_l^\dagger\alpha_l\vert0>_M$, that is, the average value in the Minkowski vacuum of the Rindler occupation number operator $n_l=\alpha_l^\dagger\alpha_l$ is now straightforward through the use of (141) and the formula $$\vert\Gamma(iy)\vert ^2={{\pi}\over{ySh\pi y}}, \ \ y\neq 0.$$ The result is $$_M<0\vert n_l\vert 0>_M=z{{c^2}\over{a}}\times{{1}\over{e^{{{\hbar\omega_l}\over{k_B{\cal T}}}}-1}}\eqno{(157)}$$ where $z=\int_{0}^{+\infty}{{dk}\over{k}}$ is an infinite constant factor (analogous to $\delta(0)$ in the previous calculation, equation (132)), and $${\cal T}={{\hbar a}\over{2\pi k_Bc}}\eqno{(158)}$$ is the Unruh temperature.

\

{\bf IV. Hawking temperature.}

\

An approximation of the Schwarzschild coordinates in the neighborhood of the event horizon of the Schwarzschild black hole, allows us to apply the Unruh effect in Rindler space to find the Hawking temperature of the black hole radiation.

\

{\it IV.1. Surface gravity of the Schwarzschild black hole}

\

The Schwarzschild metric in the Schwarzschild coordinates $(t,r,\theta,\phi)$ is given by $$ds^2=(1-{{r^*}\over{r}})c^2dt^2-{{dr^2}\over{1-{{r^*}\over{r}}}}-r^2d^2\Omega, \ \ d^2\Omega=d\theta^2+sin^2\theta d\phi^2.\eqno{(159)}$$ $r^*={{2GM}\over{c^2}}$ is the Schwarzschild radius or radius of the event horizon and $M$ is the gravitating mass. Since the metric is time independent, $$K_{(0)}=K_{(0)}^\mu{{\partial}\over{\partial x^\mu}}, \ \ K_{(0)}^\mu=\delta^\mu_0\eqno{(160)}$$ is a timelike Killing vector field for $r>r^*$: $\vert\vert K_{(0)}\vert\vert^2=\vert\vert {{1}\over{c}}{{\partial}\over{\partial t}}\vert\vert^2=1-{{r^*}\over{r}}>0$. ($[K_{(0)}]=[L]^{-1}$.) The singularity at $r=r^*$ is a coordinate singularity and is eliminated passing to another set of coordinates e.g. Eddington-Finkelstein or Kruskal-Szekeres. The only physical singularity is at $r=0$. 

\

The 4-velocity of a {\it static observer} at $r>r^*$ is $$u^\mu_{obs}(r)=(c(1-{{r^*}\over{r}})^{-1/2},\vec{0})=c(1-{{r^*}\over{r}})^{-1/2}K^\mu_{(0)}\eqno{(161)}$$ with $\vert\vert u_{obs}(r)\vert\vert^2=c^2$. We can write $$K_{(0)}=f(r)u_{obs}(r), \ \ f(r)={{1}\over{c}}(1-{{r^*}\over{r}})^{1/2}={{1}\over{c}}\sqrt{g_{00}}.\eqno{(162)}$$ Notice that $f(\infty)={{1}\over{c}}$, $f(r^*)=0$, with $f^\prime(r)={{r^*}\over{2cr^2}}(1-{{r^*}\over{r}})^{-1/2}\buildrel{r\to r^*_+}\over\longrightarrow +\infty$, $0<f(r)<{{1}\over{c}}$ for $r\in(0,\infty)$. 

\

The relation between proper time at $r$, $d\tau(r)$, and coordinate time $t$, is $$d\tau=(1-{{r^*}\over{r}})^{1/2}dt.$$ So, proper time coincides with coordinate time at $r=\infty$. The same relation holds for the period of a wave and therefore also for the wavelengths, in particular for light. Then, $$\lambda(r)=(1-{{r^*}\over{r}})^{1/2}\lambda_\infty \ \ or \ \ \lambda_\infty=(1-{{r^*}\over{r}})^{-1/2}\lambda(r).\eqno{(163)}$$ So, $\lambda_\infty\to\infty$ as $r\to r^*_+$: light emitted from the horizon has an {\it infinite red shift} at large distances. The function $(1-{{r^*}\over{r}})^{1/2}$ is called the {\it red shift factor}. One has $$\lambda(r)=cf(r)\lambda_\infty \ \ \Longleftrightarrow \ \ \nu_\infty=cf(r)\nu(r).\eqno{(164)}$$ For light emitted at the horizon, $\nu_\infty=cf(r^*)\nu(r^*)=0$. 

\

The {\it surface gravity} ($\kappa$) of a Schwarzschild black hole is the magnitude of the 4-acceleration of a static observer at $r^*$ as measured by a static observer at $r=\infty$. 

\

{\it Note}: A static observer at $r$ must be accelerated; on the contrary its motion would be geodesic i.e. in free fall.

\

Let us compute $\kappa$. The 4-acceleration of a particle or observer at $r$ is $$a^\mu={{Du^\mu}\over{d\tau}}={{dx^\nu}\over{d\tau}}{{Du^\mu}\over{dx^\nu}}={{dx^\nu}\over{d\tau}}(\partial_\nu u^\mu+\Gamma^\mu_{\nu\rho}u^\rho)={{dx^\nu}\over{d\tau}}\partial_\nu u^\mu+\Gamma^\mu_{\nu\rho}{{dx^\nu}\over{d\tau}}{{dx^\rho}\over{d\tau}}={{d^2x^\mu}\over{d\tau^2}}+\Gamma^\mu_{\nu\rho}{{dx^\nu}\over{d\tau}}{{dx^\rho}\over{d\tau}}.\eqno{(165)}$$ The (non vanishing) Christoffel symbols $\Gamma^\alpha_{\beta\gamma}={{1}\over{2}}g^{\alpha\delta}(\partial_\beta g_{\gamma\delta}+\partial_\gamma g_{\beta\delta}-\partial_\delta g_{\beta\gamma})$ of the Schwarzschild metric are given by $$\Gamma^r_{rr}=-{{r^*/r}\over{2(r-r^*)}}, \ \ \Gamma^r_{\theta\theta}=-(r-r^*), \ \ \Gamma^r_{\phi\phi}=-(r-r^*)sin^2\theta, \ \ \Gamma^\theta_{\phi\phi}=-sin\theta cos\theta, \ \ \Gamma^t_{tr}={{r^*/r}\over{2(r-r^*)}},$$ $$\Gamma^\theta_{\theta r}=1/r, \ \ \Gamma^\phi_{\phi r}=1/r, \ \ \Gamma^\phi_{\phi\theta}=cotg\theta, \ \ \Gamma^r_{tt}={{r^*}\over{2r^3}}(r-r^*).\eqno{(166)}$$ So, from (161) and (166), the 4-acceleration of an observer at rest at $r$ is $$a^\mu=(0,-{{c^2r^*}\over{2r^2}},0,0), \ \ a_\mu=g_{\mu\nu}a^\nu=(0,{{c^2r^*}\over{2r^2(1-r^*/r)}},0,0),\eqno{(167)}$$ and therefore $$a_\mu a^\mu=-({{c^2r^*}\over{2r^2}})^2{{1}\over{1-r^*/r}}\equiv -a^2\eqno{(168)}$$ with $$a(r)={{c^2r^*}\over{2r^2}}{{1}\over{(1-r^*/r)^{1/2}}}={{GM}\over{r^2}}{{1}\over{(1-r^*/r)^{1/2}}}.\eqno{(169)}$$ Notice that $a(r)\to \infty$ as $r\to r^*_+$, what means that to maintain an observer at rest at the horizon it is necessary an infinite acceleration. In other words, ``the acceleration of gravity at the horizon is infinite". Since acceleration or force leads to work, and this can be later transformed into radiation, the infinite red shift of $a(r)$ is the same as that of $\nu(r)$ i.e. $$a_\infty(r)=cf(r)a(r)={{GM}\over{r^2}}={{c^2r^*}\over{2r^2}}.\eqno{(170)}$$ Thus, the surface gravity is $$\kappa=a_\infty(r^*)={{GM}\over{{r^*}^2}}={{c^2}\over{2r^*}}={{c^4}\over{4GM}}.\eqno{(171)}$$ $\kappa$ decreases with $M$ because $\kappa\sim 1/{r^*}^2$ and $r^*$ increases with $M$. 

\

{\it IV.2. Rindler approximation and Hawking temperature} 

\

Define the radial coordinate $\rho$ through $$r-r^*={{\rho^2}\over{4r^*}}, \ \ r>r^*.\eqno{(172)}$$ Then $\rho=2\sqrt{r^*(r-r^*)}\in(0,+\infty)$, with $[\rho]=[L]$. From the definition of $\kappa$, $r^*={{c^2}\over{2\kappa}}$, and so $r=r^*+{{\rho^2}\over{4r^*}}={{c^2}\over{2\kappa}}+{{\kappa}\over{2c^2}}\rho^2={{c^4+(\kappa\rho)^2}\over{2\kappa c^2}}$; for $r\simeq r^*$ i.e. $$\rho\leq\rho_{max}<<{{c^2}\over{\kappa}},\eqno{(173)}$$ that is, in the neighborhood of the horizon, $r\cong{{c^2}\over{2\kappa}}$; then ${{1}\over{r}}\cong{{2\kappa}\over{c^2}}$ and so $1-{{r^*}\over{r}}=1-{{2\kappa c^2}\over{c^4+(\kappa\rho)^2}}{{c^2}\over{2\kappa}}={{(\kappa\rho)^2}\over{c^4+(\kappa\rho)^2}}\cong{{(\kappa\rho)^2}\over{c^4}}$. From (172), $dr={{1}\over{2r^*}}\rho d\rho={{\kappa}\over{c^2}}\rho d\rho$, $dr^2={{\kappa^2}\over{c^4}}\rho^2d\rho^2$, and for the Schwarzschild metric one has the approximation $$ds^2\cong{{(\kappa\rho)^2}\over{c^4}}c^2dt^2-d\rho^2-{{c^4}\over{4\kappa^2}}d^2\Omega.\eqno{(174)}$$ Let us study the 2-dimensional time-radial part of this metric. The change of variables $$T={{\rho}\over{c}}Sh({{\kappa t}\over{c}})\in(-\infty,+\infty), \ \ X=\rho Ch({{\kappa t}\over{c}})\in(0,+\infty)\eqno{(175)}$$ leads to $$ds^2_{1+1}={{(\kappa\rho)^2}\over{c^4}}c^2dt^2-d\rho^2=c^2dT^2-dX^2.\eqno{(176)}$$ That is, in the neighborhood of the black hole horizon the time-radial part of the metric is Minkowskian and therefore flat. 

\

Defining the dimensionless time coordinate $\lambda:={{\kappa t}\over{c}}$ ($[\lambda]=[L]^0$), $$ds^2_{1+1}=\rho^2d\lambda^2-d\rho^2,\eqno{(177)}$$ which clearly is of the Rindler form (28) (see Figure 2), except for the limited range of the coordinate $\rho$. From (175), $X^2-(cT)^2=\rho^2$ and so (choosing the right wedge)$$X=X(T)=+\sqrt{\rho^2+c^2T^2}\eqno{(178)}$$ with $X(0)=\rho$ and $X\buildrel{T\to\pm\infty}\over\longrightarrow\vert cT\vert$. Also, $${{cT}\over{X}}=Th\lambda \ \ i.e. \ \ cT=(Th\lambda)X.\eqno{(179)}$$ So, $\lambda=const.\Longrightarrow cT=const.X$. {\it Proper acceleration} $\alpha$ is defined by $$({{c^2}\over{\alpha}})^2=\rho^2=({{c^2}\over{\kappa}})^2e^{2\kappa\xi/c^2}.\eqno{(180)}$$ Then, $$X^2-(cT)^2=({{c^2}\over{\kappa}})^2e^{2\kappa\xi/c^2},\eqno{(181)}$$ which is (17) with $a=\kappa$, $t=T$, and $x=X$. $\xi$, with $[\xi]=[L]$, varies between $-\infty$ when $\rho\to 0_+$ and $\xi_{max}={{c^2}\over{\kappa}}ln({{\kappa\rho_{max}}\over{c^2}})$. Since $ln({{\kappa\rho_{max}}\over{c^2}})<0$, $\xi\leq\xi_{max}<0$. From (180), $$\alpha=\kappa e^{-\kappa\xi/c^2}.\eqno{(182)}$$ So, $\alpha\to +\infty$ for $\xi\to -\infty$ but never reaches the value $\kappa$ since, in the approximation considered, $\xi$ is always negative. Also, from (180), $d\rho=e^{\kappa\xi/c^2}d\xi$ i.e. $d\rho^2=e^{2\kappa\xi/c^2}d\xi^2$, then $$ds^2=e^{2\kappa\xi/c^2}(c^2dt^2-d\xi^2)\eqno{(183)}$$ which is the Rindler metric (24) with Rindler coordinates $(t,\xi)$ and $a=\kappa$. As shown in section {\it III.2.}, an accelerated observer ``sees" the temperature $${\cal T}={{\hbar\kappa}\over{2\pi k_Bc}}\eqno{(184)}$$ which, in the present context, is the Hawking temperature, i.e. ${\cal T}={\cal T}_{Hawk}$. 

\

We remark that, in contradistinction with the result (135), the result (184) does not involve the proper time of the accelerated observer, but the coordinates of the Rindler wedge. I.e., the acceleration appearing in the r.h.s. of (184) is not the proper acceleration of the observer, but the surface gravity of the black hole.

\

{\bf Acknowledgments.} The author thanks for hospitality to the Instituto de Ciencias de la Universidad Nacional de General Sarmiento (UNGS), Pcia. de Buenos Aires, Argentina, and to the Instituto de Astronom\'\i a y F\'\i sica del Espacio (IAFE) de la Universidad de Buenos Aires y el CONICET, Argentina, where this work was done. He also thanks Prof. Rafael Ferraro at IAFE and Lic. Lisandro Raviola (UNGS) for enlightening discussions, and the student Oscar Brauer (UNAM) for his help in the drawing of the figures. This work was partially supported by the project PAPIIT IN101711-2, DGAPA, UNAM, M\'exico.

\

{\bf References.}

\

Birrell, N. D. and Davis, P. C. W. (1982). {\it Quantum fields in curved space}, Cambridge University Press, Cambridge; pp.116-117.

\

Carroll, S. (2004). {\it Spacetime and Geometry. An Introduction to General Relativity}, Addison Wesley, San Francisco; p. 411.

\

Deser, S. and Levin, O. (1999). Mapping Hawking into Unruh thermal properties, {\it Physical Review D} {\bf 59}, 064004-1/7.

\

Hawking, S. W. (1974). Black hole explosions?, {\it Nature} {\bf 248}, 30-31.

\

Hawking, S. W. (1975). Particle Creation by Black Holes, {\it Communications in Mathematical Physics} {\bf 43}, 199-220; erratum: ibid. {\bf 46}, 206.

\

Landau, L. D. and Lifshitz, E. M. (1975). {\it The Classical Theory of Fields, Course of Theoretical Physics, Vol. 2}, Elsevier, Amsterdam; p. 24.

\

Lee, T. D. (1986). Are black holes black bodies?, {\it Nuclear Physics B} {\bf 264}, 437-486.

\

Reine, D. and Thomas, E. (2010). {\it Black Holes. An Introduction}, Imperial College Press, London; p. 45.

\

Rindler, W. (1966). Kruskal Space and the Uniformly Accelerated Frame, {\it American Journal of Physics} {\bf 34}, 1174-1178.

\

Unruh, W. G. (1976). Notes on black-hole evaporation, {\it Physical Review D} {\bf 14}, 870-892.

\

\

e-mail: socolovs@nucleares.unam.mx

\end